\documentclass[amsmath,amssymb,aps]{revtex4} 
\usepackage{graphicx} 
\usepackage{bm} 
\usepackage[usenames]{color} 
\usepackage{multirow}
\usepackage{amsmath}

\pdfoutput=1

\newcommand{\ecoli}{\emph{E. coli}}
\newcommand{\paeru}{\emph{P. aeruginosa}}
\newcommand{\llact}{\emph{L. lactis}}

\begin{document}

\title{Genome landscapes and \\
bacteriophage codon usage}

\author{Julius B. Lucks$^1$} \author{David R. Nelson$^{1,2}$}
\author{Grzegorz Kudla$^1$} \author{Joshua B. Plotkin$^{3,*}$}
\affiliation{ $^1$FAS Center for Systems Biology, Harvard University,
\\ $^2$ Lyman Laboratory of Physics, Harvard
University\\ $^3$ Department of Biology, University
of Pennsylvania\\ $^*$E-mail:
jplotkin@sas.upenn.edu }

\date{\today} 
\begin{abstract}
    
    Across all kingdoms of biological life, protein-coding genes exhibit
    unequal usage of synonmous codons. Although alternative theories abound,
    translational selection has been accepted as an important mechanism that
    shapes the patterns of codon usage in prokaryotes and simple eukaryotes.
    Here we analyze patterns of codon usage across 74 diverse bacteriophages
    that infect \emph{E. coli}, \emph{P. aeruginosa} and \emph{L. lactis} as
    their primary host. We introduce the concept of a `genome landscape,' which
    helps reveal non-trivial, long-range patterns in codon usage across a
    genome. We develop a series of randomization tests that allow us to
    interrogate the significance of one aspect of codon usage, such a GC
    content, while controlling for another aspect, such as adaptation to
    host-preferred codons. We find that 33 phage genomes exhibit highly
    non-random patterns in their GC3-content, use of host-preferred codons, or
    both. We show that the head and tail proteins of these phages 
    exhibit significant bias towards host-preferred codons, relative
    to the non-structural phage proteins. Our results support the hypothesis of
    translational selection on viral genes for host-preferred codons, over a
    broad range of bacteriophages.

\end{abstract}

\maketitle

\section{Introduction}\label{sec:introduction} 

The genomes of most organisms exhibit significant codon bias -- that is, the
unequal usage of synonymous codons. There are longstanding and contradictory
theories to account for such biases. Variation in codon usage between taxa,
particularly within mammals, is sometimes atrributed to neutral processes --
such as mutational biases during DNA replication, repair, and gene conversion
\cite{Bern95,Francino1999,Galtier2003,Eyre91}.

There are also theories for codon bias driven by selection. Some researchers
have discussed codon bias as the result of selection for regulatory function
mediated by ribosome pausing \cite{LawrHart91}, or selection against
pre-termination codons \cite{Fitc80,ModiBatt81}. However, the dominant selective
theory of codon bias in organisms ranging from \textit{E. coli} to
\textit{Drosophila} posits that preferred codons correlate with the relative
abundances of isoaccepting tRNAs, thereby increasing translational efficiency
\cite{ZuckPaul65,Ikem81a,Ikem85,PoweMori97,DebrMarz94,SoreKurl89} and accuracy
\cite{Akas94}. This theory helps to explain why codon bias is often more extreme
in highly expressed genes \cite{Ikem81b}, or at highly conserved sites within a
gene \cite{Akas94}. Translational selection may also explain variation in codon
usage between genes selectively expressed in different tissues
\cite{Plotkin2004,Dittmar2006}. However, recent work suggests that synonymous
variation, particularly with respect to GC content, affects transcriptional
processes as well \cite{Kudla2006}.

The codon usage of viruses has also received considerable attention
\cite{Jenkins2003,PlotDush03}, particularly in the case of bacteriophages
\cite{Sharp1984,Kunisawa1998,Sahu2004, Sahu2005,Sau2005,SauGosh2005}. Most work
along these lines has focused on individual phages, or on the patterns of
genomic codon usage across a handful of phages of the same host.

Here, we provide a systematic analysis of intragenomic variation in
bacteriophage codon usage, using 74 fully sequenced viruses that infect a
diverse range of bacterial hosts. Motivated by energy landscapes associated with
DNA unzipping \cite{LubenskyNelson2002,Weeks2005}, we develop a novel
methodological tool, called a genome landscape, for studying the long-range
properties of codon usage across a phage genome. We introduce a series of
randomization tests that isolate different features of codon usage from each
other, and from the amino acid sequence of encoded proteins. More than twenty of
the phages in our analysis are shown to exhibit non-random variation in
synonymous GC content, as well as non-random variation in codons adapted for
host translation, or both. Additionally, we demonstrate that phage genes
encoding structural proteins are significantly more adapted to host-preferred
codons compared to non-structural genes. We discuss our results in the context
of translational selection and lateral gene transfer amongst phages.

\section{Results}\label{sec:results} 

\subsection{Genome Landscapes}\label{sub:genome_landscapes} 

We start by introducing the concept of a genome landscape, which provides a
simple means for visualizing long-range correlations of sequence properties
across a genome. A genome landscape is simply a cumulative sum of a specified
quantitative property of codons. The calculation of the cumulative sum is
straightforward, and it consists of scanning over the genome sequence one codon
at a time, gathering the property of each codon, and summing it with the
properties of previous codons in the genome sequence. 
Similar cumulative
sums are used in solid-state physics for, e.g., the
the calculation of energy levels 
\cite{Ashcroft1976}. 
In the case of the GC3 landscape, we have
\begin{equation} 
    \label{eq:FGC3}
    F_{\mathrm{GC3}}(m) = \sum_{i=1}^m
(\eta_{\mathrm{GC3}}(m) - \overline{\eta_{\mathrm{GC3}}}) 
\end{equation}
where $\eta_{\mathrm{GC3}}(m)$ equals one or zero, depending upon whether the
the $m^{th}$ codon ends in a G/C or A/T, respectively. Note that we subtract the
genome-wide average GC3 content, $\overline{\eta_{\mathrm{GC3}}}$, so that
$F_{\mathrm{GC3}}(0) = F_{\mathrm{GC3}}(N) = 0$, where $N$ is the length of the
genome. In other words, we convert the genome codon sequence into a binary
string of 1's and 0's according to whether each codon is of type GC3 or AT3, and
we cumulatively sum this sequence to compute $F_{\mathrm{GC3}}(m)$.

The interpretation of a GC3 landscape is straightforward. Regions of the genome
whose landscape exhibits an uphill slope contain higher than average GC3
content, whereas regions of downhill slope contain lower than average GC3
content. The genome landscape provides an efficient visualization of long-range
correlations in sequence properties across a genome, similar to the techniques
introduced by Karlin \cite{Karlin1993}.

\begin{figure}
	[p] 
	\begin{center}
		\begin{tabular}
			{cc} 
			\includegraphics[scale=0.8]{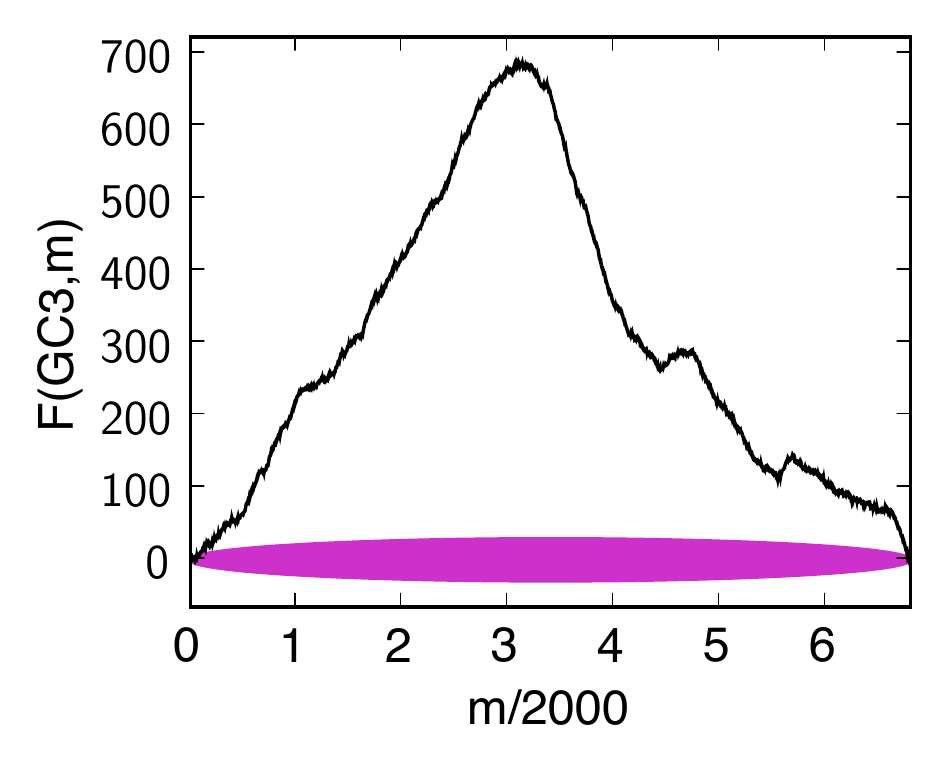} & 
			\includegraphics[scale=0.8]{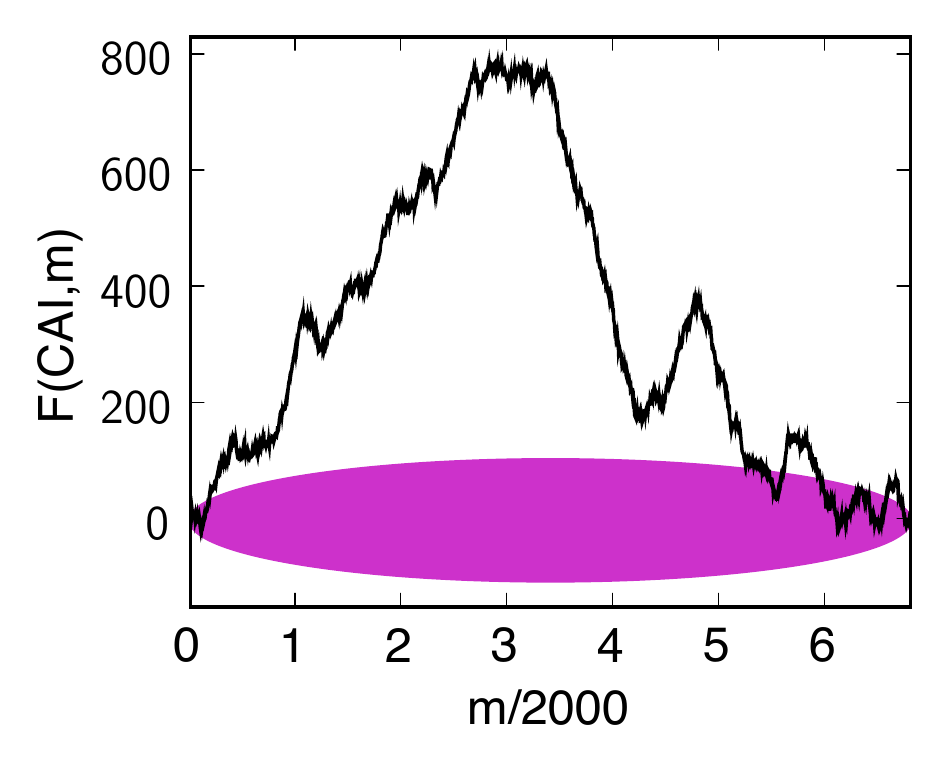} \\
			\includegraphics[scale=0.8]{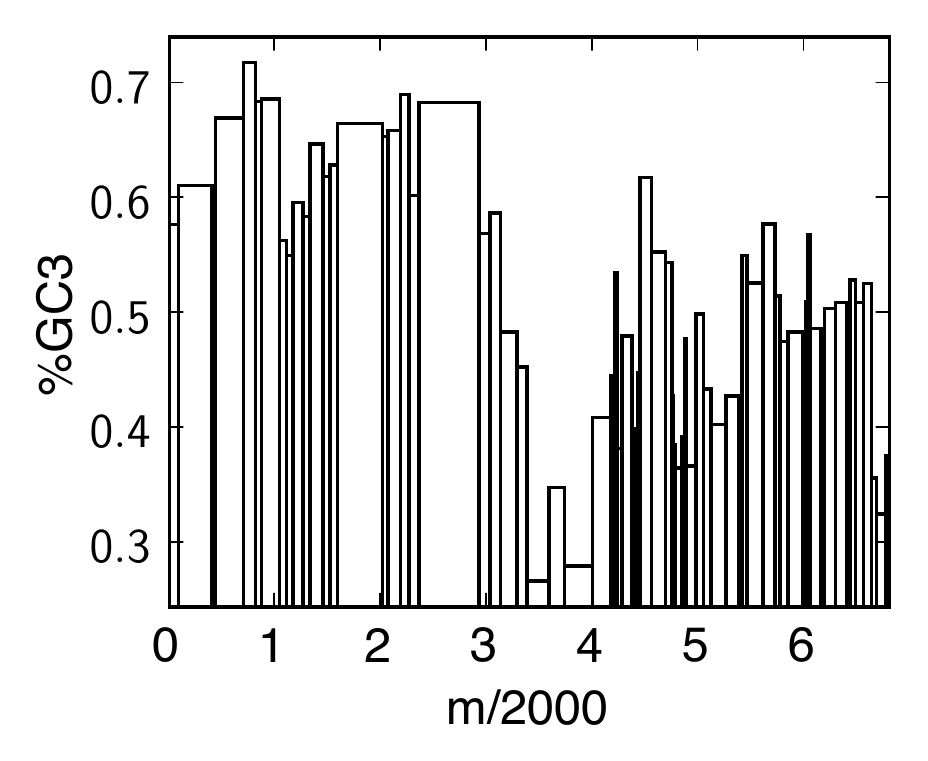} & 
			\includegraphics[scale=0.8]{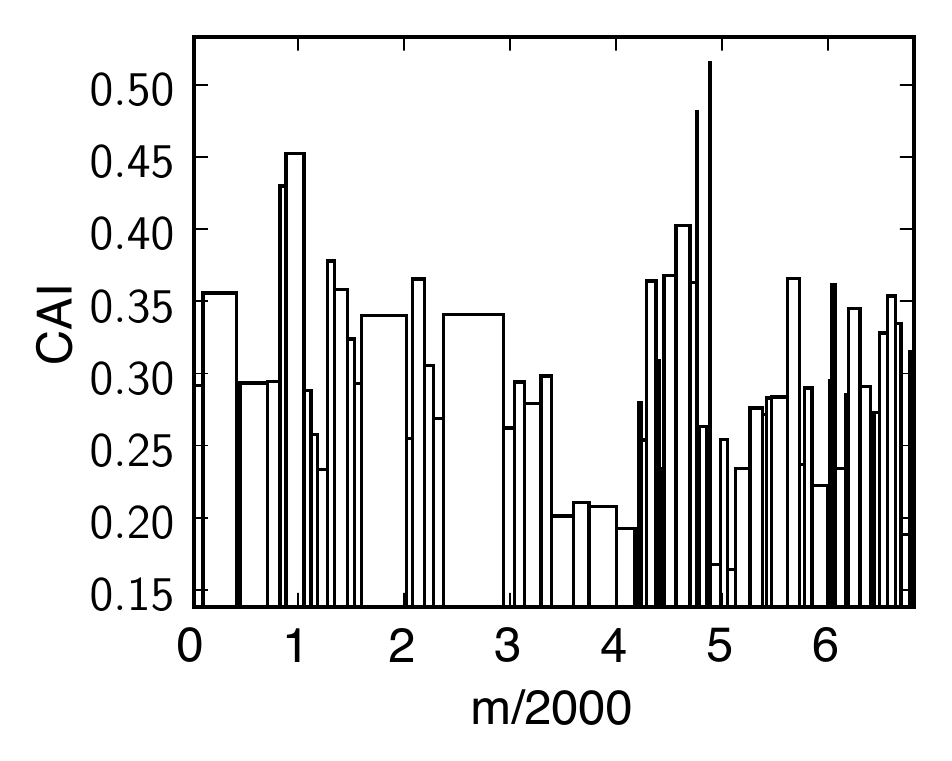} \\
		\end{tabular}
	\end{center}
	\caption{{\bf GC3 and CAI landscapes for lambda phage.} Landscapes of GC3
(left) and CAI (right) measures of codon usage in Lambda phage. Only
coding sequences are considered, which when concatenated together are 40,773 bp
long (see Table \ref{tab:phage_properties}). The GC3 landscape is the
mean-centered cumulative sum of the GC3 content (GC3=1, AT3=0) of codons. The
CAI landscape is the mean-centered cumulative sum of the log $w$-value for each
codon. For each landscape, a region
exhibiting an uphill slope corresponds to higher than average GC3 or CAI. The
horizontal purple band represents the expected
amount of variation in a random walk of GC3 or AT3 choices, 
given by Equation \eqref{eq:sigma}. Both landscapes exhibit features far
outside of the purple bands, indicating that the patterns of codon
usage are highly non-random. Gene boundaries are represented by the bars in the
histograms below each landcape. The height of the bars in the histogram indicate
the GC3 and CAI values for each gene.} \label{fig:land_hist}
\end{figure}

Traditional visualizations of GC3 content involve moving window averages of
\%GC3 over the genome \cite{Gregory2006}. In order to compare these techniques
with the landscape approach, we focus on the \emph{E. coli} phage lambda as an
illustrative example. Figure \ref{fig:land_hist} (a) shows the lambda phage GC3
landscape above its associated ``GC3 histogram". The histogram shows the GC3
content of each gene, and the width of each histogram bar reflects the length of
the corresponding gene. The figure reveals a striking pattern of lambda phage
codon usage: the genome is apparently divided into two halves that contain
significantly different GC3 contents \cite{Inman1966,Sanger1982}. The large
region of uphill slope on the left half of the GC3 landscape reflects the fact
that the majority of the genes in this region contain an excess of codons that
end in G or C. This trend is also reflected in the GC3 histogram bars, which are
higher than average in the left half of the genome (Figure \ref{fig:land_hist}).

Genome landscapes also provide a natural means of evaluating whether or not
features of codon usage are due to random chance. Under a null model in
which
the $\eta(i)$'s above are chosen as independent random variables
with $\mathrm{var}(\eta(i)) = \langle \eta(i)^2 \rangle
- \langle \eta(i) \rangle^2 = \Delta$, one can show (see
Methods) that the standard
deviation of $F(\mathrm{GC3},m)$ is
\begin{equation}
    \label{eq:sigma}
    \sigma_{\mathrm{GC3}}(m) = \sqrt{\langle
    F(\mathrm{GC3},m)^2 \rangle - \langle F(\mathrm{GC3},m) \rangle^2} = \sqrt{\frac{\Delta_{\mathrm{GC3}} m (N-m)}{N}}.    
\end{equation}
This quantity is shown as a purple band in Figure
\ref{fig:land_hist}. For $\eta(i)$'s chosen to be 0 or 1 at random, 
$\Delta_{\mathrm{GC3}} = 1/4$ and the maximum width $\sqrt{N}/4$ is
obtained at $m= N/2$. Since the scale of variation across the lambda phage GC3
landscape is much greater than its expectation under the null, we can
conclude that the distribution of G/C versus A/T ending codons is
highly non-random in the lambda phage genome.

\begin{figure}
	[p] 
	\begin{center}
		\begin{tabular}
			{cc} 
			(a) \includegraphics[scale=0.8]{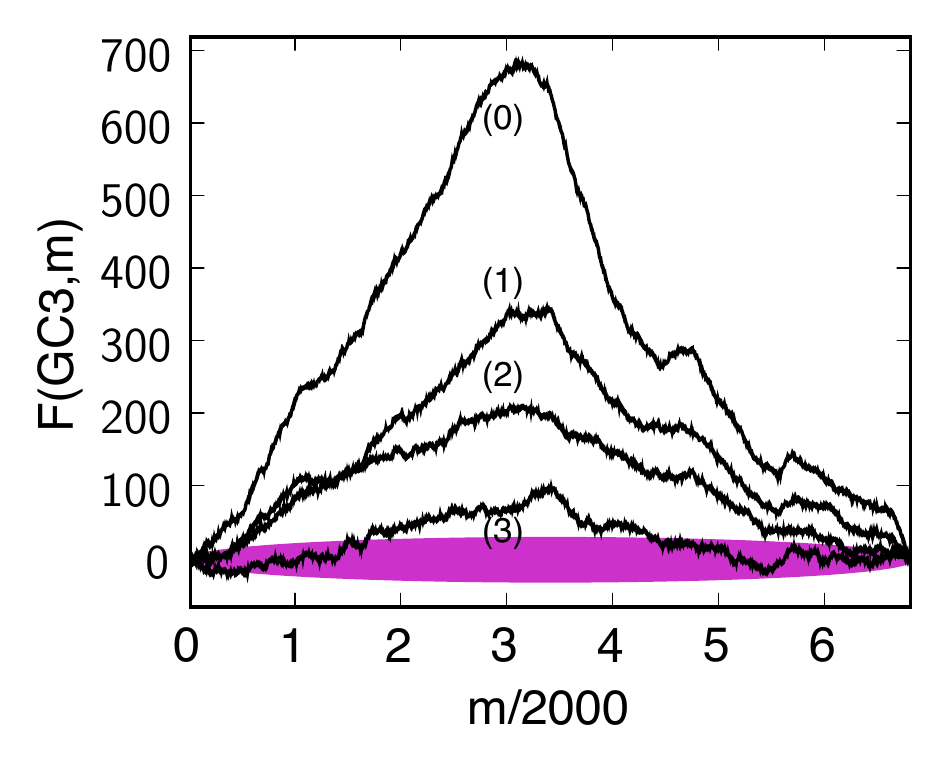} & 
			(b) \includegraphics[scale=0.8]{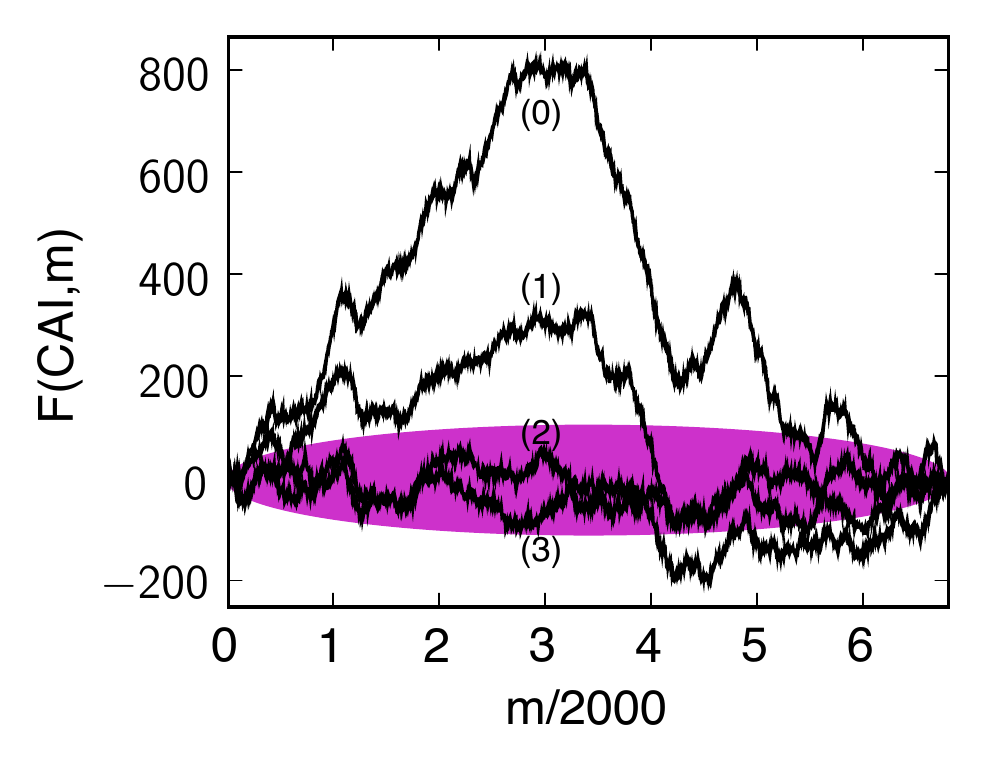} \\
		\end{tabular}
	\end{center}
	\caption{ {\bf Snapshots of simulated 
synonymous mutation in the lambda phage genome.} Panel (a) shows GC3 and
(b) shows CAI landscapes. In between successive snapshots (labeled 
by integers), $N$ synonymous mutations are introduced into the genome and the
resulting landscape is shown, where $N$ is the number of codons in the lambda
phage genome (see Section \ref{sub:genome_landscapes}). These snapshots show that
the simulated genome landscapes approach the random null model, indicated by
the purple band (see Figure \ref{fig:land_hist}). The final CAI landscape (3) lies
almost completely within the purple band. Using the lambda phage mutation rate
of $7.7\mathrm{x}10^{-8}$ mutations/bp/replication \cite{Drake1991}, we can
estimate that approximately $10^7$ genome replications would be
required
to relax within the purple bars.} \label{fig:land_decay}
\end{figure}

We can also gain intuition about the degree of non-randomness in the GC3
landscape by considering what would happen if the lambda phage genome were to
accumulate random synonymous mutations.  Figure
\ref{fig:land_decay}(a) shows snapshots of the lambda GC3 landscape as
we simulate synonymous mutations to the genome. Between each snapshot,
$N$ synonymous mutations were introduced by
picking a codon at random along the genome, and then choosing a new
synonymous codon at random according to the global lambda phage codon
distribution. As more mutations are introduced, the GC3 landscape of the
synonymously mutated lambda genome approaches the purple band,
indicating that the GC3 pattern in the real lambda phage genome is
highly non-random.

The procedure of producing a genome landscape can be applied to other
properties of codon usage. In addition to GC3, we will study patterns in
the Codon Adaptation Index (CAI).  CAI measures the similarity of a
gene's codon usage to the `preferred' codons of an organism
\cite{Sharp1987} -- in this case, the host bacterium of the phage under
study.  Every bacterium has a preferred set of codons defined as the
codons, one for each amino acid, that occur most frequently in genes
that are translated at high abundance. These genes are often taken to be
the ribosomal proteins and translational elongation factors
\cite{Sharp1987} (see Methods).

In order to calculate CAI, the preferred codons are each assigned a weight $w =
1$. The remaining codons are assigned weights according to their frequency in
the highly-translated genes, relative to the frequency of the $w=1$ codon. The
CAI of a gene is defined as the geometric mean of the $w$-values for its codons
\begin{equation}
\label{eq:CAI_def} \mathrm{CAI} = \left(\Pi_{i=1}^{M} w_i\right)^{1/M},
\end{equation} 
where $w_i$ is the $w$-value of the $i^{th}$ codon, and
$M$ is the length of the gene. This quantity can be re-written as
\begin{equation} \mathrm{CAI} = \exp(\frac{1}{M} \sum_{i=1}^{M}
\ln(w_i)).  
\end{equation} 
The latter formulation is more useful for calculating genome landscapes,
because the argument of the exponential function is now a sum of the logs of the
$w$-values. Therefore, we define the CAI landscape as \begin{equation}
F_{\mathrm{CAI}}(m) = \sum_{i=1}^m (\eta_{\mathrm{CAI}}(m) -
\overline{\eta_{\mathrm{CAI}}}), \end{equation} where $\eta_{\mathrm{CAI}}(m) =
\ln(w_m)$.

The CAI landscape for lambda phage is shown in Figure
\ref{fig:land_hist}(b), along with the CAI histogram of lambda phage.
For the CAI histograms, the height of each bar represents the CAI value
of that gene (Eq. \ref{eq:CAI_def}).  As in the case with the GC3
landscape, we find that the lambda phage CAI landscape corresponds
closely to the CAI histogram, but it offers a more striking global view
of the long-range CAI structure in the lambda phage genome. One
contiguous half of the lambda phage genome exhibits elevated CAI,
whereas the other half exhibits depressed CAI.  The observed CAI
landscape lies far outside the purple band in Figure
\ref{fig:land_hist}, calculated according to Eq. \ref{eq:sigma},
indicating that the pattern of CAI across the lambda phage genome is
non-random. However,
the purple band is wider for the CAI landscape than for the GC3
landscape, because the variance in the $\ln{(w_i)}$'s,
$\Delta_{\mathrm{CAI}}$, is greater than $\Delta_{\mathrm{GC3}}$.

The GC3 and CAI landscapes for lambda phage are highly correlated with each
other (Figure \ref{fig:land_hist}). In particular they both have large uphill
regions on the left-hand side of the genome, indicating a region containing
codons with elevated GC3-content and CAI values, compared to the genome average.
It is possible that the observed correlation between the GC3 and CAI landscapes
could be caused by the conflation between high CAI and GC3 in the preferred
\emph{E. coli} codons, as we discuss below.

We note that the genes in the region of elevated CAI primarily encode the highly
translated structural proteins that form the capsid and tail of the lambda phage
virions. This patterns suggests the hypothesis that, because of the need to
produce structural genes in high copy number during the viral life cycle, structural
genes preferentially use codons that match the host's preferred set of codons.
We will explore this translational-selection hypothesis in greater detail below.

\subsection{The Effect of Amino Acid Content on Genome
Landscapes}\label{sub:the_effect_of_amino_acid_content_on_genome_landscapes} 

The previous section illustrated that the codon usage across the lambda phage
genome is highly non-random with respect to both GC3 and CAI. In this section we
quantify this statement, and we focus on aspects of lambda's codon usage
patterns that are \emph{independent} of the amino acid sequences of the
encoded proteins.

Since we are interested in studying the patterns of \emph{synonymous} codon
usage, it is important that we control for the amino acid sequence of encoded
proteins. Phages utilize a diverse spectrum of proteins, ranging from
those that form the protective capsid for nascent progeny, to those
encoding for the tail and tail fibers, to those that regulate the switch
between lytic or lysogenic infection pathways. As with other organisms, phage
proteins have been selected at the amino acid level for function and folding.
Some portion of a phage's codon usage is surely influenced by selection
for amino acid content.

We can construct a simple randomization test to interrogate the potential
influence of the amino acid sequence on the GC3 and CAI landscapes of lambda
phage. In this test, we generate random genomes that have the exact same amino
acid sequence as lambda phage, but shuffled codons, such that the genome-wide,
or global, codon distribution is preserved in each random genome (see
Methods). As
summarized in Table \ref{tab:tests}, we refer to this test as the `aqua'
randomization test. For each of the randomized genomes, we calculate GC3 and CAI
landscape. Similar to a recent randomization method \cite{Zeldovich2007}, we then 
compare the observed landscape of the actual genome to the
distribution of landscapes generated from the randomized genomes.

\begin{figure}
	[p] 
	\begin{center}
		\begin{tabular}
			{cc} 
			\includegraphics[scale=0.8]{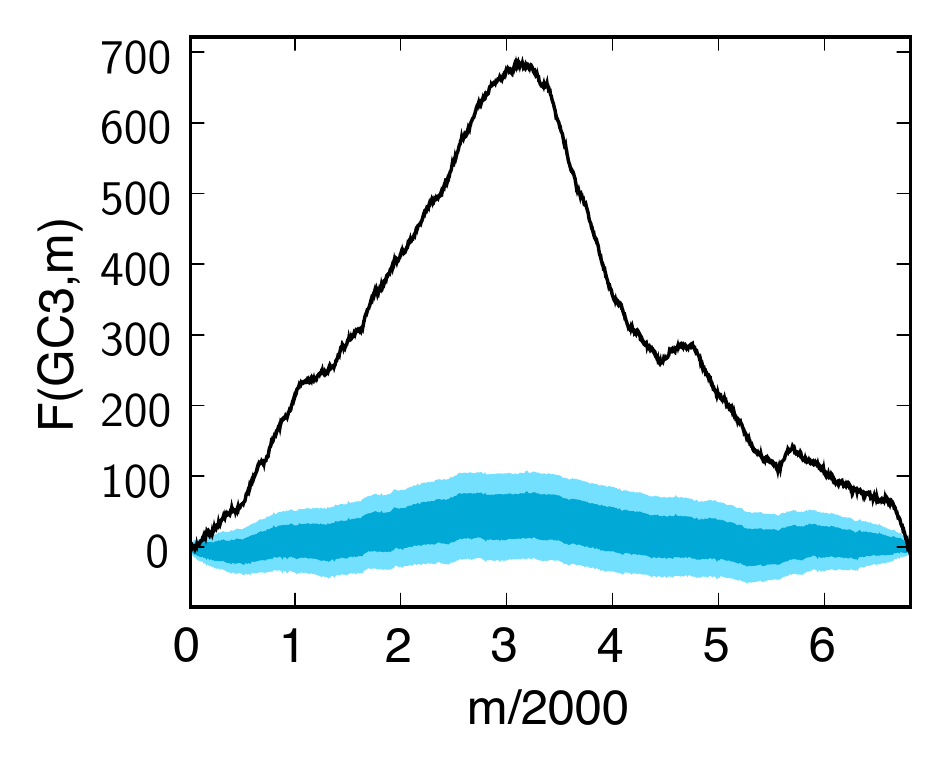} & 
			\includegraphics[scale=0.8]{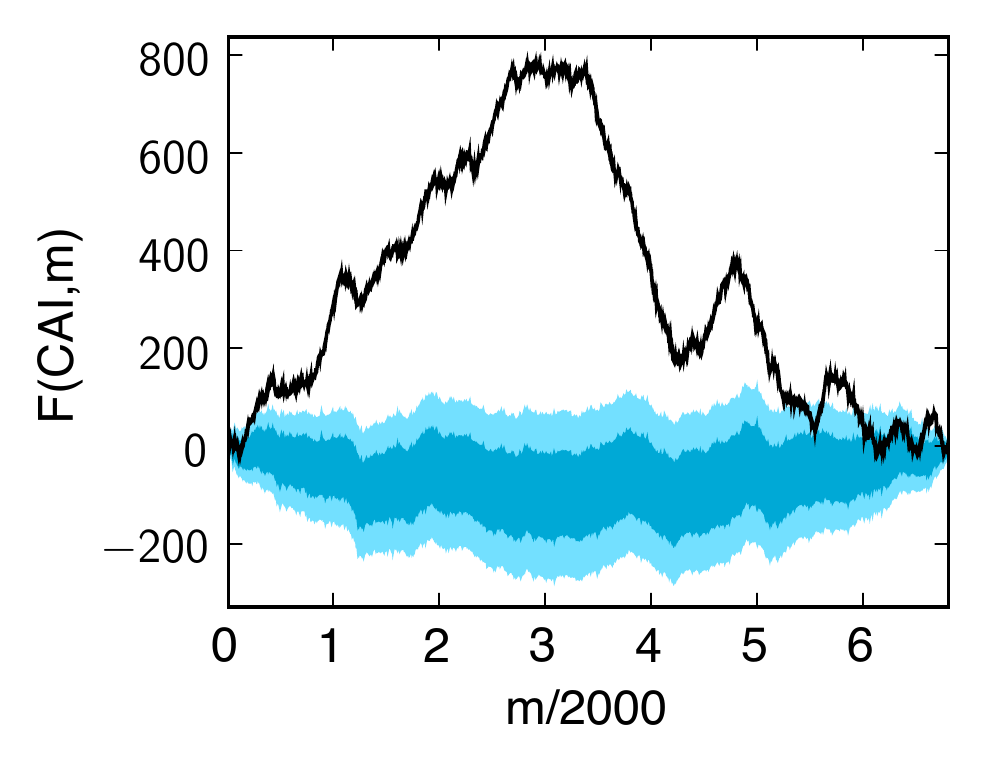} \\
			\includegraphics[scale=0.8]{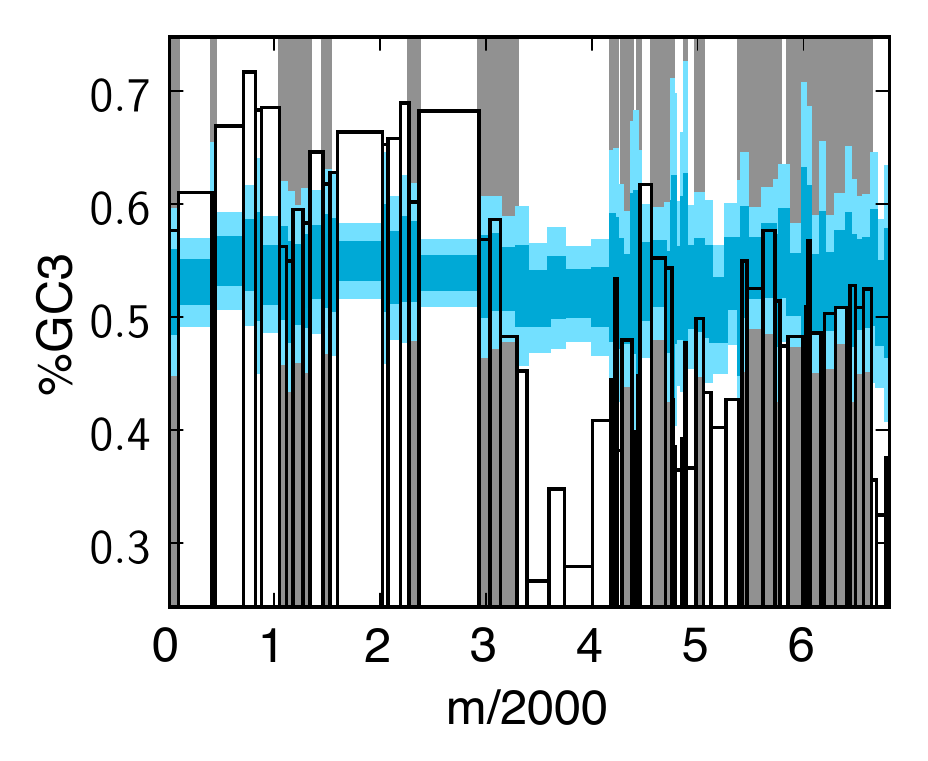} & 
			\includegraphics[scale=0.8]{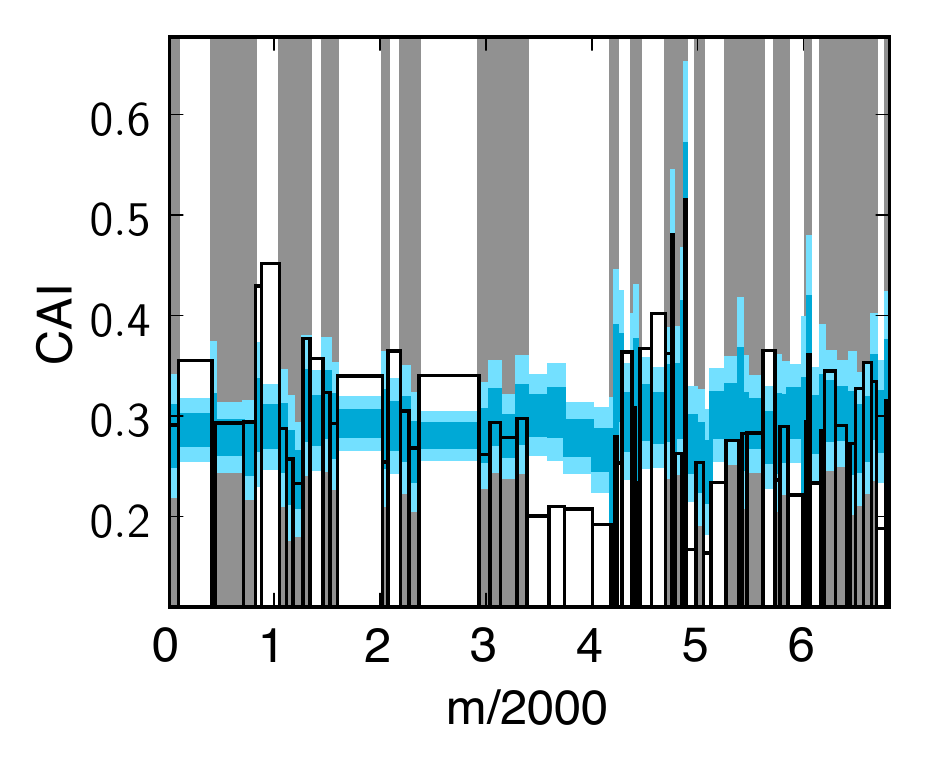} \\
		\end{tabular}
	\end{center}
	\caption{{\bf Observed and randomized landscapes for lambda phage. } The figure
shows the observed GC3 (left) and CAI (right) landscapes, plotted in black,
along with the mean $\pm 1$, and $\pm 2$ standard deviations of
randomized trials, shown in aqua (bold line, dark and light regions,
respectively). The `aqua' randomization test shown here draws random
synonymous codons that preserve the exact amino acid
sequence, according to probabilities that preserve the global codon
usage
distribution of the lambda genome. For the most part, the observed landscapes
lie signficantly outside the distribution of randomized landscapes -- implying
that the amino acid content of genes is not responsible for the observed pattern
of the CAI landscape. In the lower panel, however, genes whose GC3 (left), or
CAI (right) values fall between the 0.025 and 0.975 quantile of the random
trials are shadowed in grey; the GC3/CAI values of such genes are not
significantly different from random, given their amino acid sequence.}
\label{fig:aqua}
\end{figure}

Figure \ref{fig:aqua} shows the results of this comparison, with the observed
landscapes plotted as black lines, and the mean
$\pm$ one and two standard deviations of random trials shown in dark and light
aqua, respectively. As the figures show, the observed landscapes lie in the far
extremes of the randomized distributions -- indicating that the amino acid
sequence of the lambda phage genome does not determine the extraordinary
features of the observed landscapes.

It is also instructive to query the influence of amino acid content on codon
usage in each gene individually. The histogram view of these randomization tests
allows us to ask this question precisely. Because the amino acid sequence is
preserved exactly across the genome, each histogram bar in Figure \ref{fig:aqua}
can be considered as its own randomization test, one for each gene. The position
of the horizontal black bar reflects the actual codon usage of
each gene, and it can be compared to the distribution of random trials in order
to compute a quantile for each gene: 
\begin{equation} q^{>} = \frac{\mathrm{number\ of\
trials\ less\ than\ observed}}{\mathrm{number\ of\ trials}},\\
q^{<} = \frac{\mathrm{number\ of\ trials\ greater\ than\
observed}}{\mathrm{number\ of\ trials}}.  
\end{equation} 
Note that we have defined two quantiles, $q^{>}$ and $q^{<}$, that describe the
proportion of random trials strictly less or strictly greater than the observed data.
These two quantities sum to a values less than one (and equal to one if there
are no ties). A large value of $q^{>}$ signifies that the observed statistic
(e.g. GC3 or CAI) is \emph{greater} than most of the random trials.

Associated with each of these quantiles is a p-value quantifying whether the
observed gene sequence has significantly different codon usage than the random
trials: $p^{<} = 1 - q^{<}$ and $p^{>} = 1 - q^{>}$. If either one of these
$p$-values is low, it signifies that the GC3 (or CAI) content of the gene is
significantly different than the genomic average, controlling for the amino acid
sequence of the gene. $p^<$ tests for significantly depressed GC3 (or CAI) in a
gene; and $p^>$ tests for significantly elevated GC3 (or CAI) in a gene. We will
use these $p$-values, which arise from the `aqua' randomization test, in two
ways.

Since we are interested in studying the effects of synonymous codon usage alone,
we first wish to filter out any genes whose codon usage does not significantly deviate
from random, given the amino acid sequence. Therefore, in the subsequent
gene-by-gene analyses reported in this paper, we retain only those genes whose
quantiles fall in the extreme 5\% of random trials. That is, we only keep those
genes for which $p^{<}_{\mathrm{aqua}} < 0.025$ or $p^{>}_{\mathrm{aqua}} <
0.025$. These genes are said to `pass' the aqua test, and they are
unshaded in Figure \ref{fig:aqua}.

We also use the gene-by-gene $p$-values to quantify the degree to which
codon usage is independent of amino acid sequence across the genome as a
whole. To do so, we combine all the gene-by-gene $p$-values into an
aggregate $p$-value for the entire genome, $p_{\mathrm{aqua}}$, using
the method of Fisher \cite{Fisher1948}. We calculate the combined
$p$-value by summing the logs of twice the minimum of each gene-specific
p-value \begin{equation} f_{\mathrm{aqua}} = -2 \sum_{i=1}^{i=k} \ln{[2
\min(p^{<}_{\mathrm{aqua},i}, p^{>}_{\mathrm{aqua},i})]}, \end{equation}
where $p^{<}_{\mathrm{aqua},i}$ represents the aqua $p^<$-value for gene
$i$, and $k$ is the number of genes in the genome. It is well known that
$f_{\mathrm{aqua}}$ is chi-squared distributed with $2k$ degrees of
freedom \cite{Fisher1948}.  Thus, the combined $p$-value for the
entire genome, $p_{\text{combined}}^{\mathrm{aqua}} = 1-
P_{\chi^2,2k}(f_{\mathrm{aqua}})$, where $P_{\chi^2,2k}(f)$ is the
cumulative chi-squared distribution with $2k$ degrees of freedom. In
the case of lambda phage, we find $p_{\text{combined}}^{\mathrm{aqua}} =
7.42\mathrm{x}10^{-98}$ for GC3 and $p_{\text{combined}}^{\mathrm{aqua}}
= 1.50\mathrm{x}10^{-41}$ for CAI. Thus, we conclude that the neither
the GC3 nor the CAI patterns across the lambda phage genome are
determined by the genome's amino acid sequence.

In the following sections we will use the aqua test (see Table
\ref{tab:tests}) and its associated gene-by-gene and combined p-values
as a control to verify that features of codon usage are not driven by
the amino acid sequence.

\subsection{Disentangling CAI from GC3}\label{sub:disentangling_cai_from_gc3} 

Depending upon the preferred codons of the host species, the effect of
selection for high CAI in a viral gene is not necessarily independent
from the effect of selection for other features of viral codon usage,
such as high GC3.
For example, codons with high CAI values associated with a given host
may be biased towards high GC3 values as well (see Figure
\ref{fig:E_coli_master}, and Section
\ref{sub:disentangling_cai_from_gc3} below). It is important, therefore,
to disentangle the effects of selection for CAI versus selection for
GC3, in order to determine which one of these forces is responsible for
the non-random patterns of codon usage observed in the lambda genome.

The weights used to compute CAI for \emph{E. coli} are shown in Figure
\ref{fig:E_coli_master}. The 61 codons are placed into one of four groups
according to whether they are GC3 or not (red or blue, respectively), and
whether they have high CAI or not (dark or light, respectively). High CAI is
determined by an arbitrary cutoff of $w \geq 0.9$. As this table demonstrates,
the set of preferred codons in E. coli is slightly biased towards GC-ending
codons (58\%).

The GC bias of preferred codons, although slight, could conflate the
results of selection for CAI versus GC3 in phages that infect \emph{E.
coli}, such as lambda.  We therefore introduce another randomization
test that allows us to disentangle patterns of CAI content from patterns
of GC3 content. Similar to the aqua randomization test described above,
we draw random phage genomes such that the amino acid sequence is
conserved, but we add the additional constraint of conserving the exact
GC3 sequence as well (see Methods). For example, at a site containing a
GC3 codon for leucine, in our random trials we only allow those leucine
codons terminating in G or C. By comparing the observed landscapes of
the genome with the distribution of randomly drawn landscapes, we can
isolate the features of codon usage driven by CAI, independent of GC3
and amino acid content. We refer to this randomization procedure at the
`orange' randomization test (Table \ref{tab:tests}).

Conversely, we also wish to assess the strength of patterns in GC3 content,
independent of CAI and amino acid content. The appropriate randomization
procedure in this case requires that we constrain the amino acid sequence and
the sequence of codon CAI values while allowing GC3 to vary. However, because
CAI values are not binary, CAI cannot be constrained exactly while still
allowing for enough variability to produce a meaningful randomization test.
Thus, we introduce a binary version of the CAI measure, called BCAI, that is
qualitatively the same as and, for our purposes, interchangeable with CAI.

\begin{figure}
	[p] 
	\begin{center}
		\begin{tabular}
			{cc} 
			\includegraphics[scale=0.8]{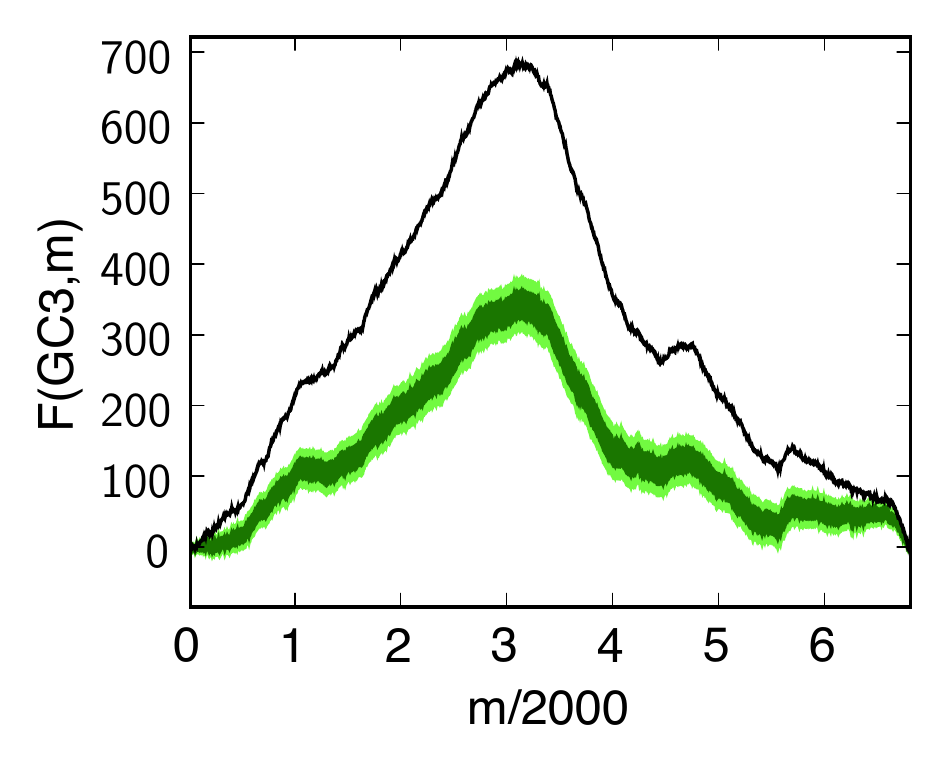} & 
			\includegraphics[scale=0.8]{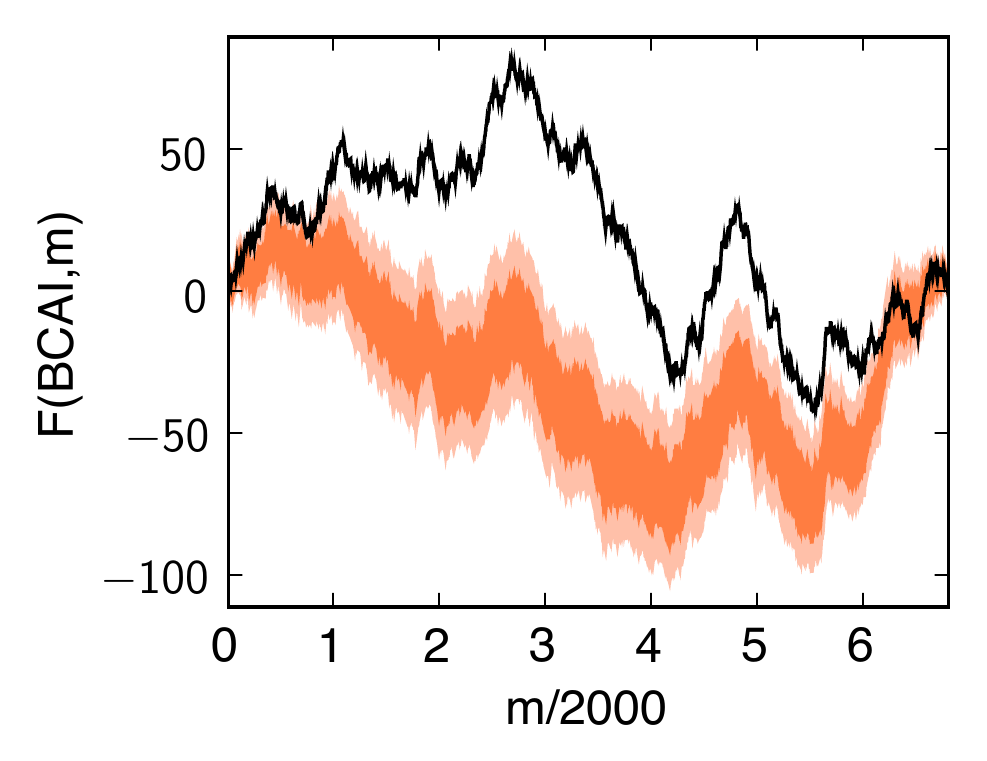} \\
			\includegraphics[scale=0.8]{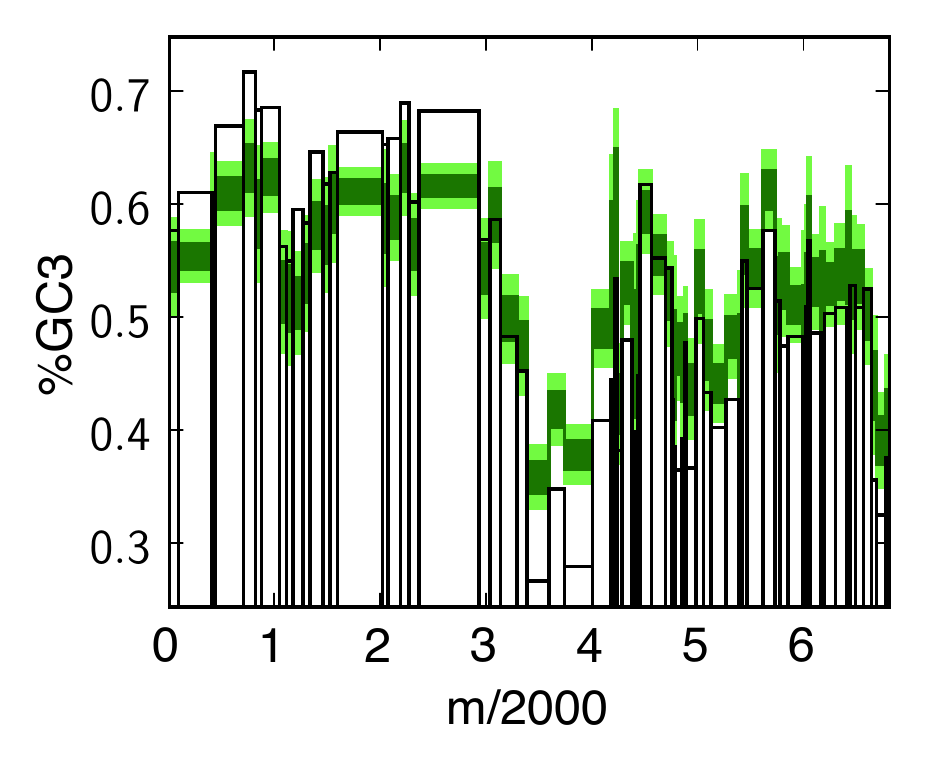} & 
			\includegraphics[scale=0.8]{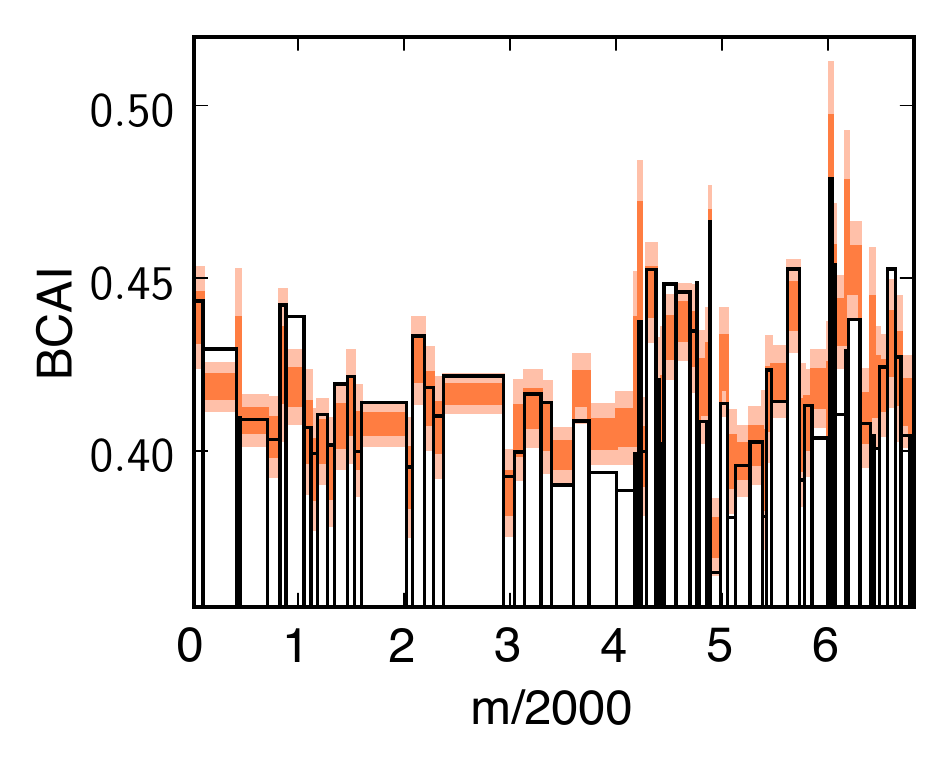} \\
		\end{tabular}
	\end{center}
	\caption{{\bf Observed and randomized landscapes for lambda phage.} 
	Observed landscapes are shown along with randomized landscapes 
	associated with the `green' and `orange' tests. 
	The green randomization procedure tests the 
	significance of the GC3 landscape controlling for the observed 
	CAI (actually, BCAI) variation across the genome. The orange 
	randomization procedure tests the significance of the BCAI landscape, 
	controlling for the observed GC3 variation across the genome. 
	Both tests preserve the amino-acid sequence exactly. 
	Both observed landscapes lie outside the distribution 
	of random trials, indicating there is non-random GC3 
	content controlling for CAI, and non-random CAI 
	content controlling for GC3.} \label{fig:green_orange} 
\end{figure}

The BCAI $w$-value for a codon is defined to be 0.7 if the codon is high CAI,
and 0.3 if the codon has low CAI. High CAI is defined by the threshold of $w
\geq 0.9$ (see Figure \ref{fig:E_coli_master}). The actual values assigned for
BCAI are arbitrary and have no effect on our results. In addition, the threshold
value $w \geq 0.9$ is also arbitrary, and our results are robust to changing
this threshold. BCAI provides a useful surrogate for CAI because its values are
binary, thereby allowing us to constrain a gene's amino acid sequence and BCAI
sequence \emph{exactly}, while varying GC3 content in random trials. The BCAI
landscapes and histograms are calculated in the same way as CAI landscapes and
histograms, except using BCAI $w$-values. As expected, the BCAI landscape of a
genome is qualitatively similar to its CAI landscape (compare Figures
\ref{fig:green_orange}b and \ref{fig:aqua}b), and the two landscapes are highly
correlated (e.g. $r = 0.72$ for lambda phage). Thus BCAI is interchangeable
with CAI for the purposes of our randomization tests.

Figure \ref{fig:green_orange} shows the results of the two randomization tests
outlined above: the `green' test that compares the observed GC3 landscape to a
distribution of random trials constraining the amino acid sequence and the BCAI
sequence; and the `orange' test that compares the observed BCAI landscape to a
distribution of random trials constraining the amino acid sequence and the GC3
sequence. Our convention for naming these two tests is summarized in Table
\ref{tab:tests}.

As seen in Figure \ref{fig:green_orange}a, the observed GC3 landscape lies
significantly outside of the random trials that preserve amino acid sequence and
BCAI sequence. Combining the gene-by-gene p-values for this test, we find
$p_{\text{combined}}^{\text{green}} = 5.1\mathrm{x}10^{-68}$ -- indicating that
the lambda phage genome as a whole has non-random GC3 variation independent of
amino acid and CAI (actually, BCAI) sequence. Conversely, Figure
\ref{fig:green_orange}b shows that the BCAI landscape contains non-random
features when controlling for both GC3 and amino acid sequence
($p_{\text{combined}}^{\text{orange}} = 6.3\mathrm{x}10^{-9}$). In other words,
the lambda phage genome exhibits highly non-random patterns of both GC3 and CAI
codon variation, independent of one another and independent of the amino acid
sequence.

\subsection{Non-random patterns of CAI and GC3 In
Bacteriophages}\label{sub:selection_for_cai_and_gc3_in_bacteriophages} 


In the sections above we have demonstrated and quantified highly
non-random patterns of GC3 and CAI codon usage variation across the
lambda phage genome. We have also demonstrated that these trends are
independent of one another.  In this section, we will extend our
analysis to a large range of diverse phages.

In this section we consider all sequenced phages that infect \emph{E. coli},
\emph{Pseudomonas aeruginosa} or \emph{Lactococcus lactis} as their primary host.
The latter two hosts were chosen because of they contain unusually extreme GC3
content: 88 \%GC3 for \emph{P. aeurginosa} and 25 \%GC3 for \emph{L. lactis},
genome-wide. The extreme GC3 content of these hosts give rise to opposing
relationships between high CAI and GC3 -- as indicated schematically in
Figure \ref{fig:master_cartoons}. In particular, \emph{P. aeruginosa} strongly
favors GC3 in high-CAI codons (94\%), and \emph{L. lactis} strongly favors AT3 in
high-CAI codons (72\%). Thus, these three hosts span a large spectrum of
relationships between CAI and GC3. Since our randomization tests constrain amino
acid and BCAI exactly (the `green' test), and amino acids and GC3 exactly (the
`orange' test), we can control for any possible conflation between GC3 and CAI
trends. Thus, the randomization tests are equally applicable to all of the phage
genomes, regardless of their host.

We performed the aqua, green, and orange randomization tests on the 45
phages of \emph{E. coli}, 12 phages of \emph{P. aeruginosa}, and 17
phages of \emph{L. lactis} whose genomes have been sequenced
(see Methods). In the first step of our
analysis, we removed any phages which failed either the aqua GC3 or aqua
CAI tests, because the codon usage of such genomes are influenced by
their amino acid sequence. A phage was said to pass these two control
tests if its Fisher combined p-values for both aqua GC3 and aqua CAI
were significant. The significance criterion for each test is
$p_{\text{combined}} < 5\%/74$, which incorporates a Bonferroni
correction for multiple tests.  With this cutoff, 50 of the initial 74
phages passed the aqua control tests.

\begin{figure}
	[p] 
	\begin{center}
		\begin{tabular}
			{cc} 
		(a) \includegraphics[scale=0.5]{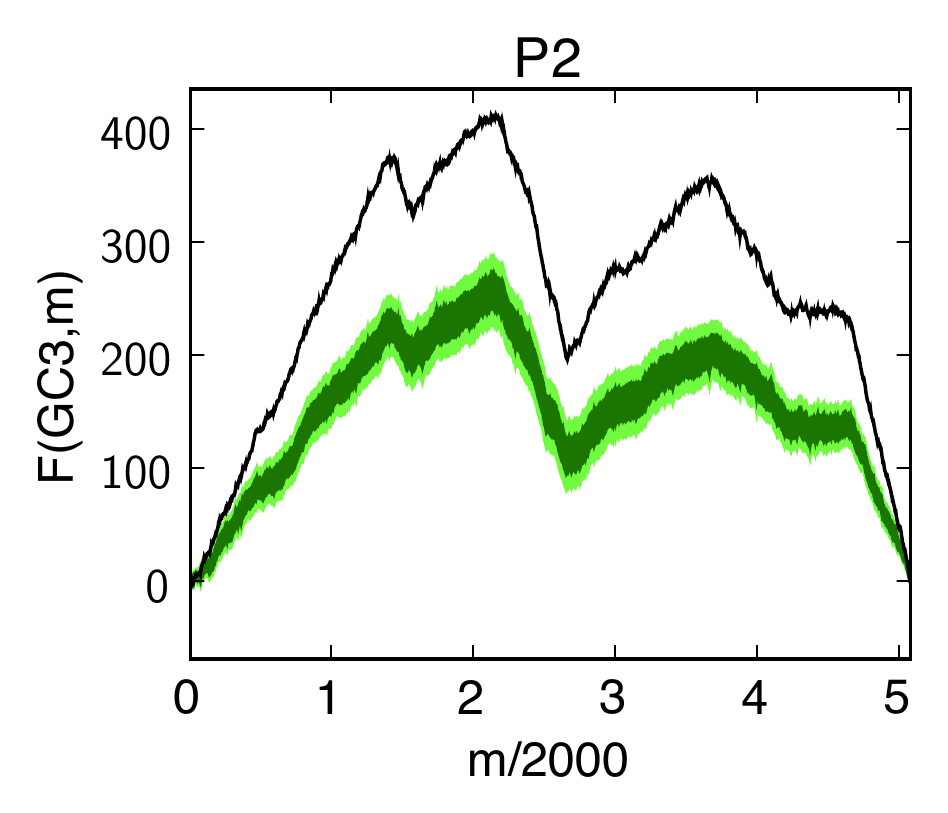} & 
			\includegraphics[scale=0.5]{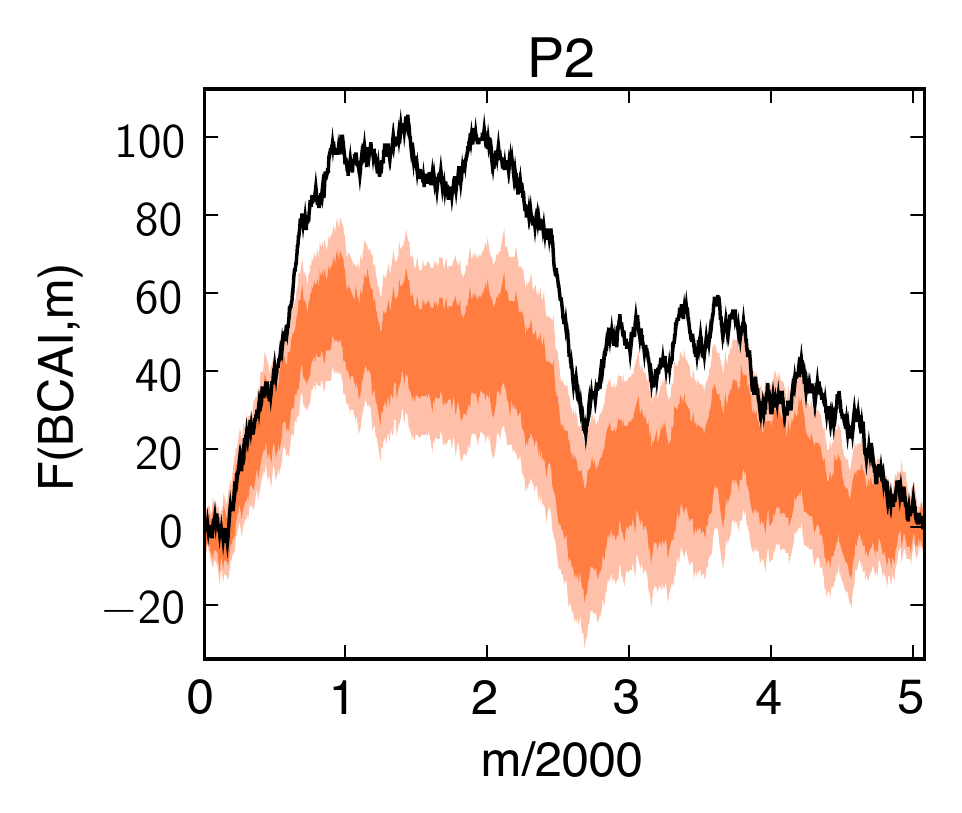} \\
		(b)	\includegraphics[scale=0.5]{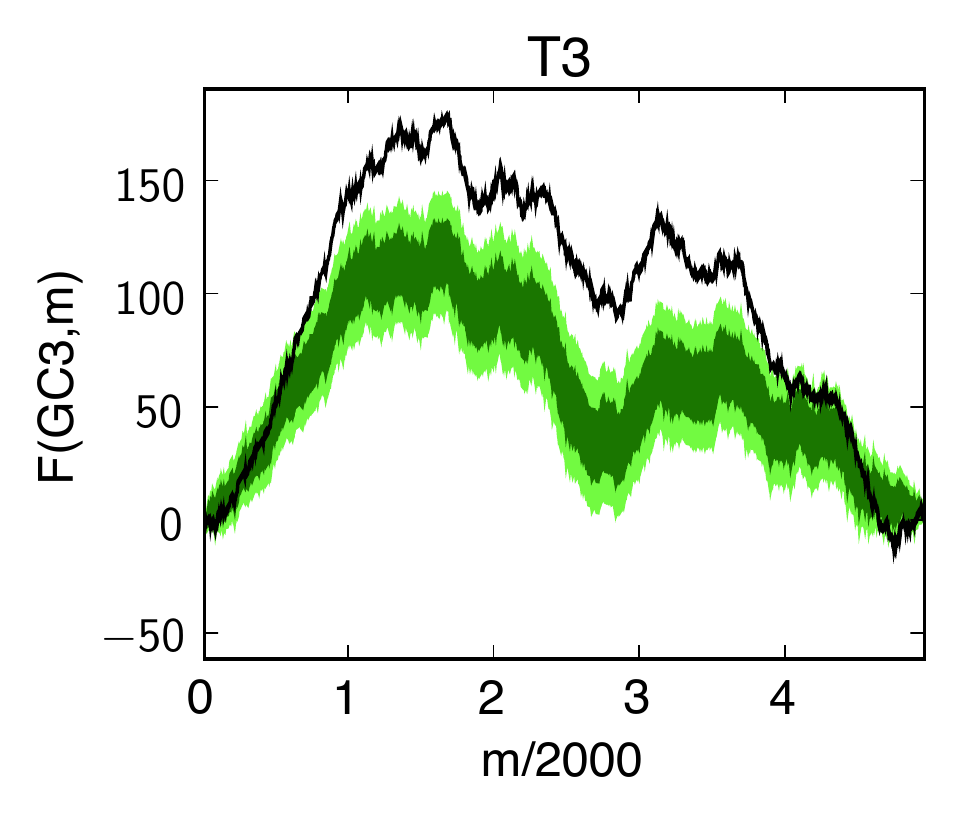} & 
			\includegraphics[scale=0.5]{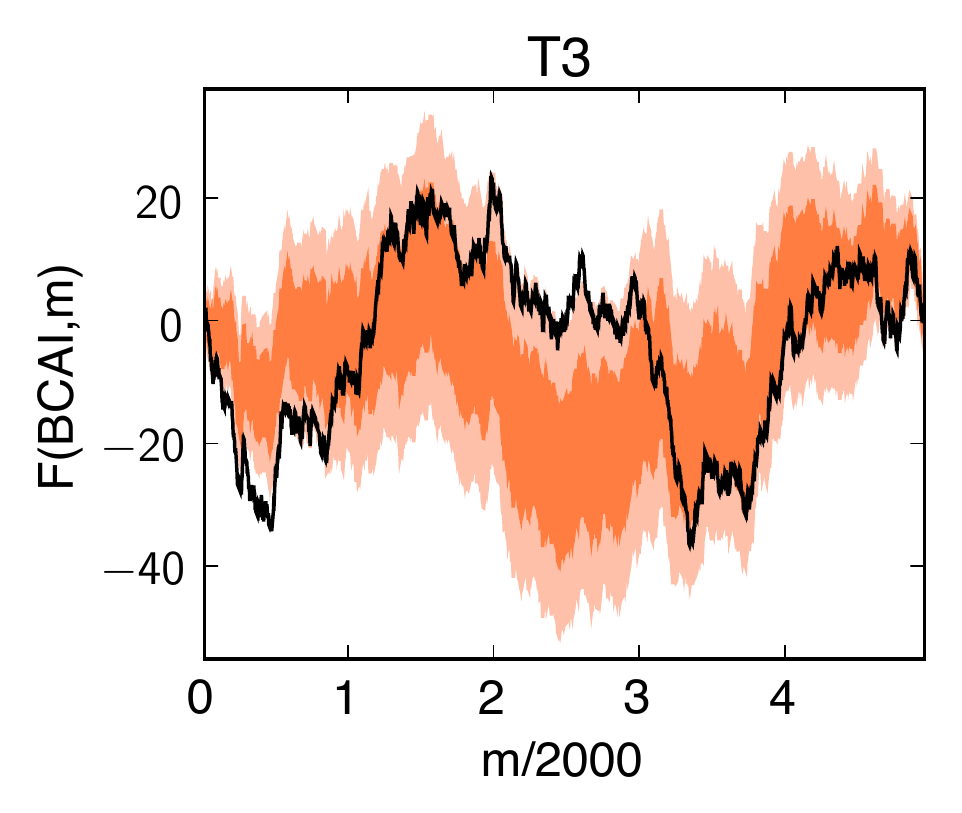} \\
		(c)	\includegraphics[scale=0.5]{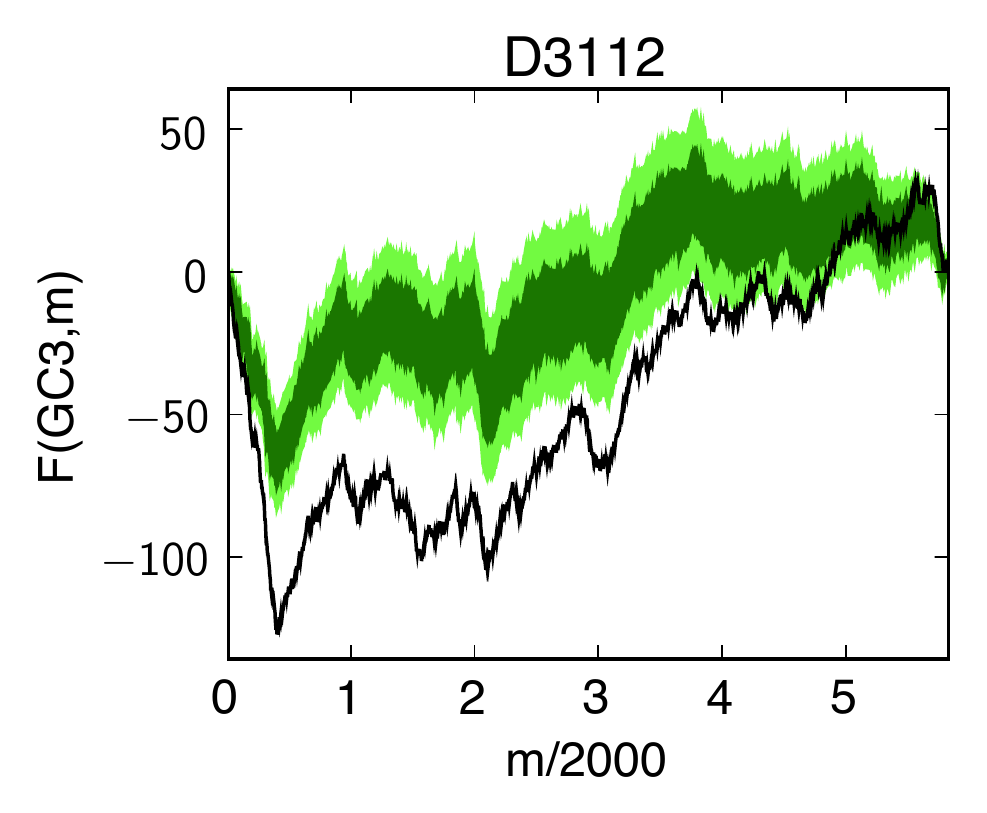} & 
			\includegraphics[scale=0.5]{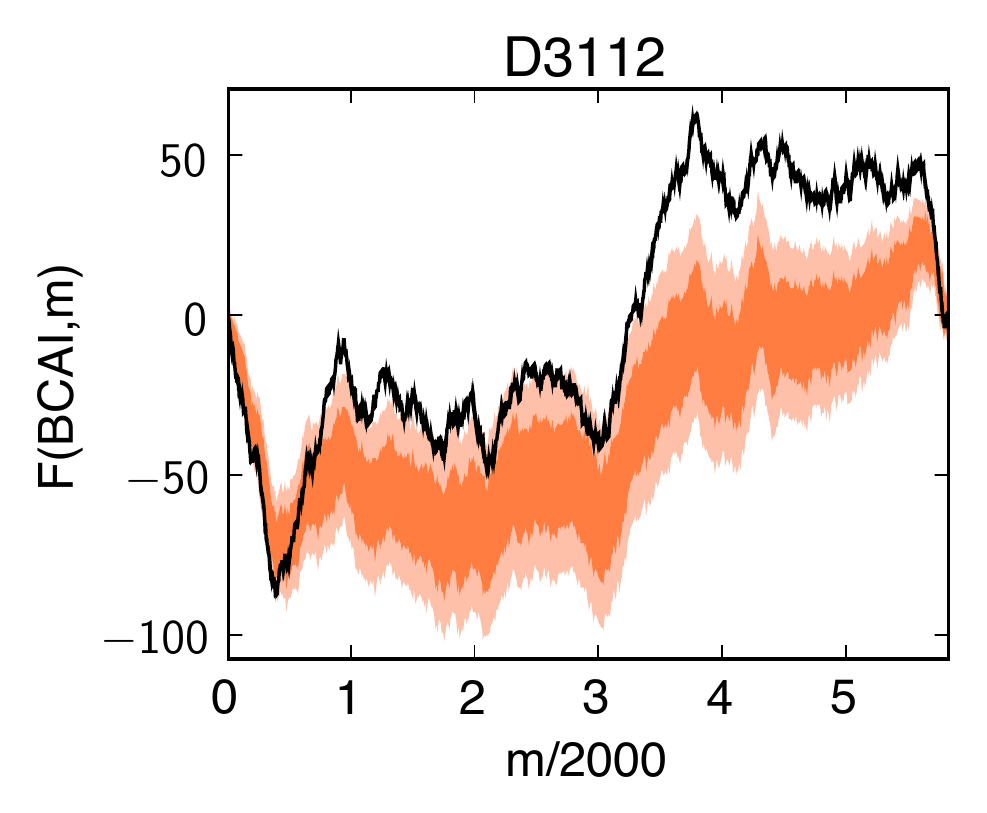} \\
		(d)	\includegraphics[scale=0.5]{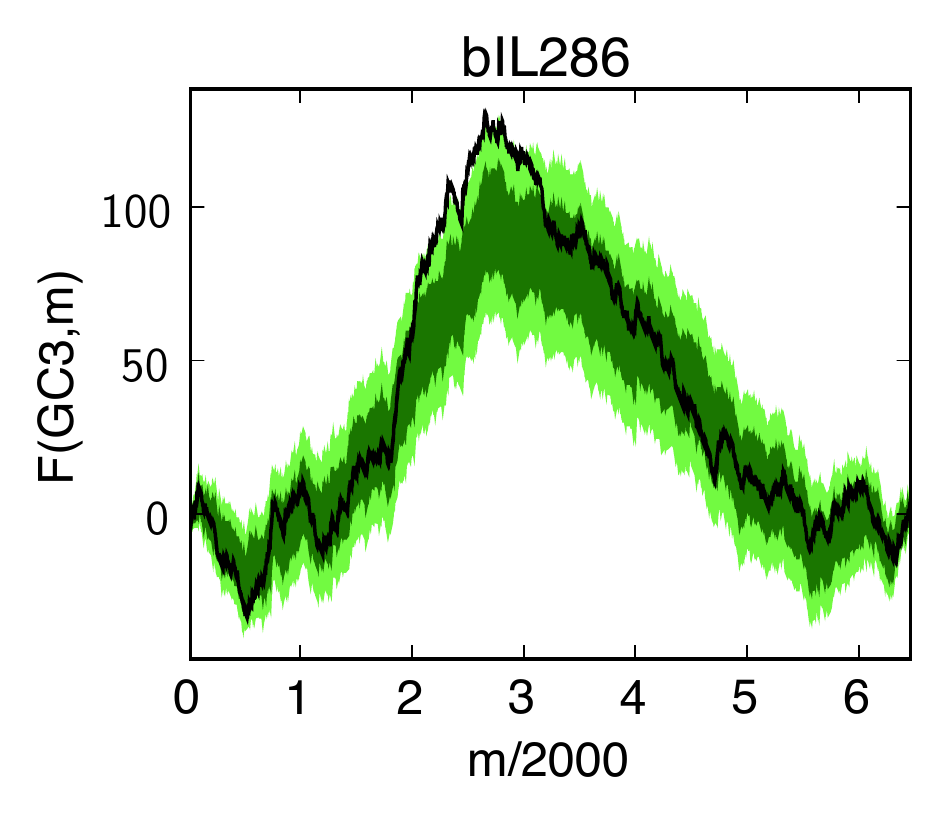} & 
			\includegraphics[scale=0.5]{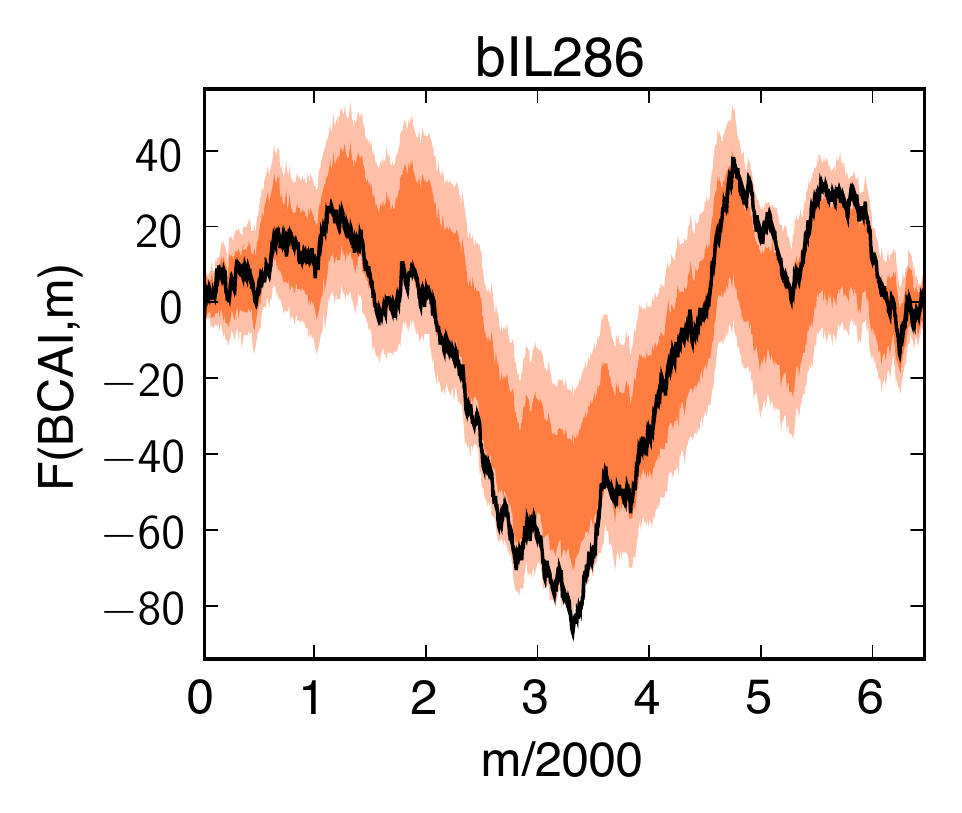} \\
		\end{tabular}
	\end{center}
	\caption{{\bf `Green' (left) and `orange' (right) randomization tests 
for several phages.} Bacteriophages P2 (b) and T3 (b) both
infect \emph{E. coli}. Phage D3112 (c) infects \emph{P. aeruginosa}.
Phage bIL286 (d) 
infects \emph{L.
lactis}. T3 is the only non-temperate phage of this group. See
Table \ref{tab:phage_properties} for combined Fisher p-values for these tests.
In the case of bIL286, note the lack of evidence for codon bias evident in 
the green and orange
tests for bIL286, as confirmed by the insignificant $p$-values in 
Table \ref{tab:phage_properties}. In this case, we cannot rule out the
possibility that the observed pattern in GC3 is determined
completely by the amino acid and CAI sequence (green), or that the observed pattern in
CAI is determined by the amino acid and GC3 sequence (orange).}
\label{fig:green_orange_examples}
\end{figure}

Figure \ref{fig:green_orange_examples} shows results of these tests for
several example genomes. P2, a temperate phage, and T3, a non-temperate
phage both infect \emph{E. coli} and both pass the control tests and
exhibit significant `orange' and `green' results, as does D3112, a
temperate phage that infects \emph{P.  aeruginosa}. However, not all
phages that pass the control test exhibit signifanct `orange' and
`green' results -- as evidenced by bIL286, a temperate phage infecting
\emph{L. lactis}.

\begin{figure}
	[p] 
	\begin{center}
		\includegraphics[scale=1]{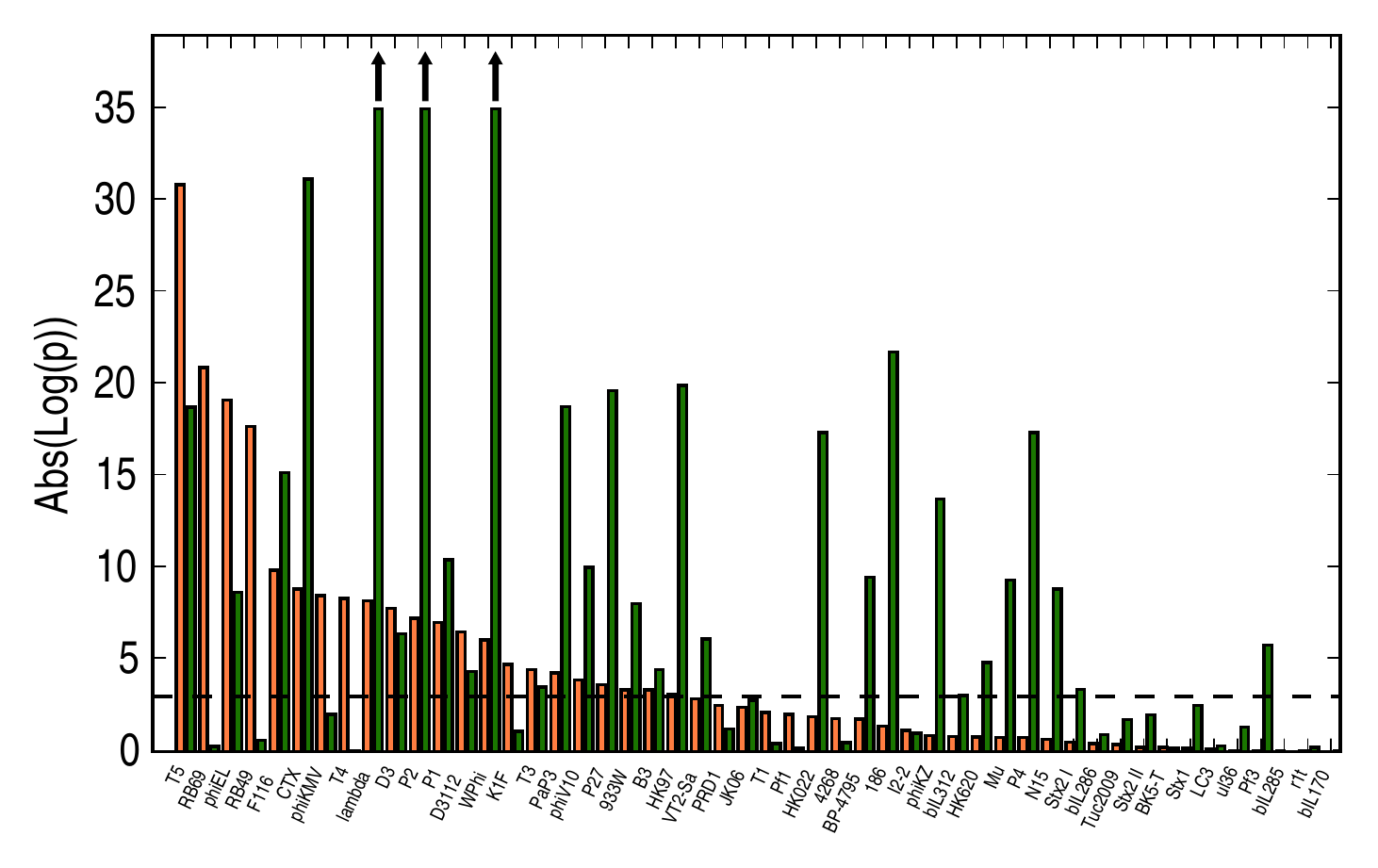} 
	\end{center}
	\caption{{\bf Combined Fisher p-values for the `green' and `orange' 
	randomization tests across 50 phage genomes.} Phage names are 
	listed on the x-axis, and are sorted by their `orange' p-value. 
	A total of 29 genomes exhibit non-random 
	GC3 content controlling for CAI (green test); and a total of 
	22 genome exhibit non-random 
	CAI content controlling for GC3 (orange test). 17 genomes pass both of 
	these tests. The dashed horizontal line indicates the 
	threshold for significance after Bonfernni correction (i.e. 5\%/50). 
	Upwards arrows indicate p-values that lie beyond the limits of the
	y-axis. See Table \ref{tab:phage_properties} for phage properties, 
	including the
	p-values for these tests. Twenty four phage genomes 
	that failed the aqua GC3 or CAI control tests 
	are not included in this figure.} \label{fig:green_orange_pass_genomes} 
\end{figure}

Figure \ref{fig:green_orange_pass_genomes} plots the distribution of
combined Fisher p-values of the orange and green tests, for the 50
phages that pass the control tests. The majority of these
p-values are highly significant. Using a Bonferoni-corrected theshold of
5\%/50, a total of 22 genomes show significance in the orange test, 29
in the green text, and 17 in both orange and green.  These results
indicate that non-random patterns in codon usage are not unique to
lambda phage.  Indeed, over a range of bacterial hosts and a range of
phage viruses, there is apparent pressure for non-random patterns of
both GC3 content and CAI content, independent of one another and
independent of the amino acid sequence.

\subsection{Translational selection on phage structural
proteins}\label{sub:translational_selection_on_phage_structural_proteins} 

In this section, we investigate a natural hypothesis concerning the patterns of
non-random CAI usage we have observed in phage genomes -- namely, that these
patterns may be driven by selection for translational accuracy and efficiency,
which is stronger in more highly expressed proteins \cite{Ikem81a,Sharp1984}. 

Among all phage proteins, the structural proteins are the most highly expressed
\cite{Hendrix2004}. The structural proteins form the protective capsid that
encloses the viral genome, as well as the tail, which is often used for
transmission of the phage genome to the inside of the host \cite{Roessner1983}.
These proteins must be produced in high copy number -- many tens of copies of
each type of structural protein needed to form each of hundreds of viral progeny
\cite{Hendrix2004}. For each gene in a phage genome, we assigned a structural
annotation of 1 if the gene was known to encode a structural protein and 0
otherwise (see Methods).

According to the standard hypothesis of translational selection, the
structural genes of phages should exhibit elevated CAI levels compared
to other phage genes, since they are translated (by the host) in high
copy numbers. To test this hypothesis, we performed regressions between
the structural annotation of phage genes and their aqua CAI and orange
BCAI p-values.  In other words, we compared the structural properties of
genes against their CAI content, controlling for amino acid sequence,
and against their BCAI content, controlling for both amino acid sequence and
GC3 sequence.

\begin{figure}
	[p] 
	\begin{center}
		\begin{tabular}
			{cc} 
			\includegraphics[scale=0.8]{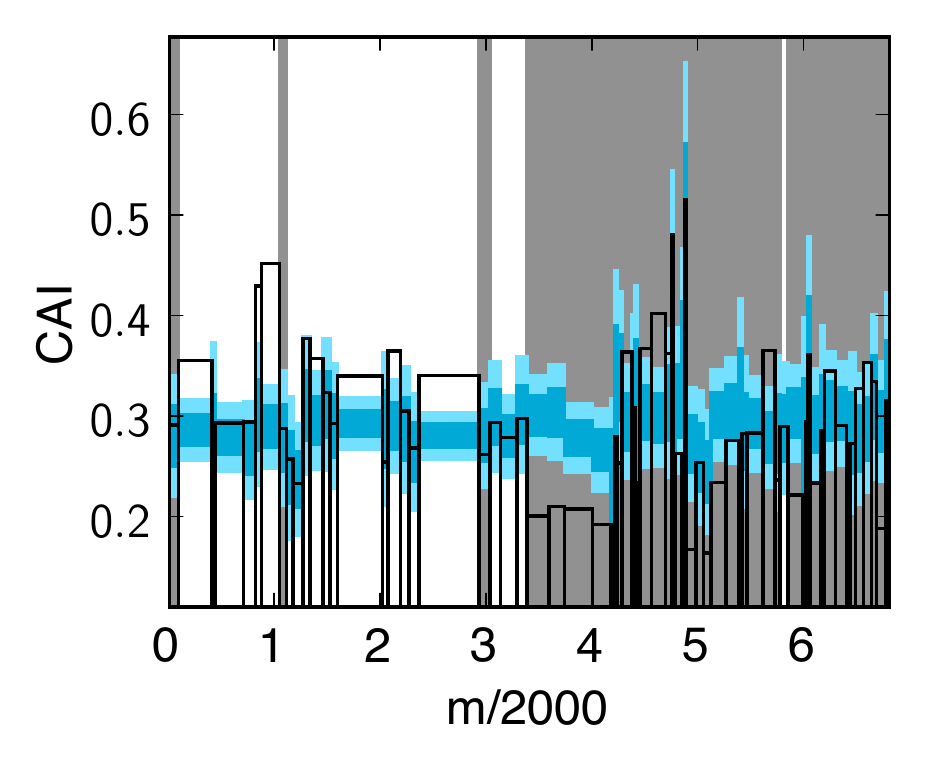} & 
			\includegraphics[scale=0.8]{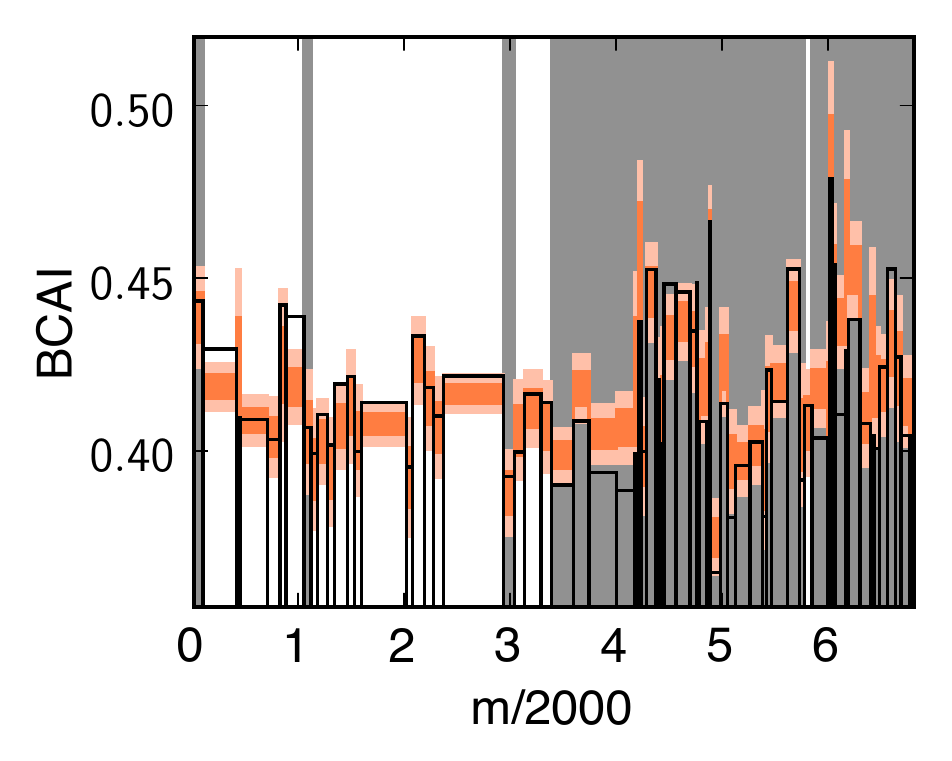} \\
		\end{tabular}
	\end{center}
	\caption{{\bf The relationship between codon usage and protein 
	function in lambda phage.} The figure shows the aqua 
	(CAI, as in Figure \ref{fig:aqua}) and orange (BCAI, as in 
	Figure \ref{fig:green_orange}) randomization tests 
	overlaid with information about protein function: 
	genes classified as structural are shown with a white background 
	and all other genes with a grey background. The histograms 
	indicate a clear relationship between the structural 
	classification of a gene and its significance under the aqua 
	and orange tests: structural genes typically have elevated 
	quantiles in the aqua test, whereas other genes typically have 
	depressed quantiles. In other words, structural genes 
	exhibit elevated CAI values when controlling for their 
	amino acid sequence, compared to codon usage in the 
	genome as a whole. Moreover, as the orange histograms 
	indicate, this pattern is not caused by variation in GC3 content: 
	the structural genes exhibit elevated BCAI values after 
	controlling for both their amino acid sequence and their 
	GC3 sequence.} \label{fig:structural} 
\end{figure}

In the case of lambda phage, Figure \ref{fig:structural} shows the results of
the aqua CAI and orange BCAI randomization tests, with the structural genes
highlighted. The plot reveals a striking pattern: the vast majority of the
structural proteins lie on the left half of the genome, exactly in the region
where genes have elevated CAI values. In order to quantify this association we
performed ANOVAs. Before regressing structural
annotations against codon usage, we first removed the non-informative genes --
i.e. genes whose codon usage are influenced by their amino acid content, as
indicated by a failure to pass the aqua CAI test.

Table \ref{tab:lambda_all_struct_non_aqua_orange} shows the results of the
regression between aqua CAI and orange BCAI $p^{>}$-values versus structural
annotations in lambda phage. The results are highly significant: structural
annotations explain half of the variation in CAI, even when controlling for
genes' amino acid sequences (aqua, $r^2$=56\%) as well as GC3 seqeuences (orange
test, $r^2$=46\%). The median $p^{>}$-value among structural genes is close to
zero, whereas the median $p^{>}$-value among non-structural genes is close to
one -- indicating that structural genes exhibit significantly \emph{elevated}
CAI values. These highly significant results are consistent with the hypothesis
of translational selection on structural proteins.

In order to examine the relationship between structural annotation and CAI
across all 74 phages in our study, we performed the same ANOVA on the 1,309
informative genes (i.e. genes that pass the aqua CAI randomization test). Once
again, Table \ref{tab:lambda_all_struct_non_aqua_orange} shows a highly
significant relationship between structural annotation and CAI values,
controlling for amino acid content and GC3. Thus, the tendency toward elevated
CAI values in structural genes holds across all the phages in this study,
despite the fact that they infect a diverse range of hosts with a wide
variety of GC contents.

\section{Discussion}\label{sec:discussion} 

In this paper, we have introduced genome landscapes as a tool for visualizing
and analyzing long-range patterns of codon usage across a genome. In combination
with a series of randomization tests, we have applied this tool to study
synonymous codon usage in 74 fully sequenced phages that infect a diverse range
of bacterial hosts. Genome landscapes provide a convenient means to identify
long-range trends that are not apparent through conventional, gene-by-gene or
moving-window analyses. Using a statistical test that compares codon usage to
random trials, controlling for the amino acid sequence, we found that
we found that many of the phages studied exhibit non-random variation 
in codon usage.  However, not all of the phages exhibit non-random variation as
exemplified by phage bIL286 (Figure \ref{fig:green_orange_examples}(d)).

In light of long-standing \cite{Ikem81a} and recent \cite{Kudla2006}
literature from other organisms, we have focussed on two aspects of
phage codon usage: variation in third-position GC/AT content (GC3) and
variation in the degree of adaptation to the `preferred' codons of the
host (CAI). Almost three-quarters of the phages in our study exhibit
non-random intragenomic patterns of codon usage, even when controlling
for the amino acid sequence encoded by the genome. Almost half of such
genomes also show non-random patterns of CAI when additionally
controlling for the GC3 sequence. In other words, there is substantial
variation in CAI above and beyond what would be expected by random
chance, given the amino acid and GC3 sequences of these genomes.

We have also compared the CAI values of phage genes to their annotations
as structural or non-structural proteins. We have conclusively
demonstrated that phage genes encoding structural proteins exhibit
significantly elevated CAI values compared to the non-structural proteins
from the same genome. These results hold even when controlling for the
the amino acid sequence and GC3 sequence of genes. Our
conclusions across a diverse range of phages are consistent with
early observations on lambda's codon usage \cite{Sanger1982},
early results for T7 \cite{Sharp1984}, and with the general hypothesis
of translational selection, which predicts elevated CAI in genes
expressed at high levels \cite{Ikem81a,Ikem81b,Sharp1987}. The pattern
of elevated CAI in structural proteins is particularly striking the case
of lambda phage. It is also worth noting that we find no
significant relationship between a phage's life-history (i.e. temperate
versus non-temperate) and the degree to which its structural proteins
exhibit elevated CAI (see Table \ref{tab:temperate_non}). This
observation likely reflects the fact that at some point every phage,
regardless of its life history, must generate certain structural proteins in
high abundance -- and so it is beneficial to encode such protein using
the host's translationally preferred codons.

Our results on translational selection in phages shed light on the
nature of selection on viruses. The standard interpretation of elevated
CAI in highly expressed bacterial proteins assumes a fitness cost (per
molecule) associated with inefficient or inaccurate translation. We have
observed a similar relationship between expression level and CAI across
a diverse range of bacteriophages, which presumably do not incur a
direct energetic cost from inefficient translation by their hosts. Thus,
our results suggest that either there is an adaptive benefit (to the
virus) of elevated CAI in phage structural proteins, or that costs
incurred by the host bacterium also reduce the fitness of the virus.

In addition to our results on CAI, we have also observed non-random patterns of
GC3 variation across the genomes of many phages. These patterns are highly
significant even after controlling for potential conflating factors, such as the
amino acid sequences and CAI sequences of genes. Unlike our results on CAI,
there is no clear mechanistic hypothesis underlying the non-random patterns of
GC3 in phages. It is possible that these patterns reflect selection for
efficient transcription \cite{Kudla2006} or for mRNA secondary structure. But in
the absence of independent information on such constraints, we cannot assess the
merits of these selective hypotheses, nor rule out the possibility of variation
in mutational biases across the phage genomes. It is interesting to note
that we find these significant non-random patterns of GC3 predominantly in
temperate phages (see Table \ref{tab:temperate_non}).

Our study benefits from the number and breadth of phages we
have analyzed. Unlike previous studies, here we analyze phages whose
suspected hosts span a diverse range of bacteria, which themselves
differ in their genomic GC3 content and preferred codon choice. We have
calibrated CAI for each phage according to its primary host, and
nevertheless we find consistent relationships between CAI and viral
protein function. These results therefore conclusively extend the
classical theory of translational selection to the relationship between
viruses and their hosts.

The present study also benefits from the development of randomization tests that
isolate the patterns of variation in CAI from variation in GC content. Due to
intrinsic biases in the GC content of the preferred codons of hosts, previously
studies on codon usage in phage have conflated these two types of synonymous
variation \cite{Sahu2004, Sahu2005,Sau2005,SauGosh2005}. The mechanisms
underlying GC3 variation and CAI variation likely differ, and so it is
critically important that we have analyzed each of these features controlling
for the other one.

There is a large literature on the structure and evolution of phage genomes
which is pertinent to our analyses of phage codon usage. The genomes of phages
that infect \emph{E. coli}, \emph{L. lactis}, and \emph{Mycobacteria} are known
to be highly mosaic in structure
\cite{Juhala2000,Brussow2002,Hendrix2002,Lawrence2002,Pedulla2003,Hatfull2006}.
In other words, these genomes exhibit many similar local features that suggest
each genome was assembled from a common pool of bacteriophage genomic regions
\cite{Hendrix1999}. Recently, mosaicism was discussed in the lambdoid
phages focusing specifically on the \emph{E. coli} phages lambda, HK97 and N15
\cite{Hendrix2004}. We note that both HK97 and N15 have peaked landscape
structures like lambda, although not as pronounced, indicating that some degree
of mosaicism can be observed in genome landscapes among closely related phages.
The postulated mechanism for mosaicism is homologous and non-homologus
recombination between co-infecting phages or between a phage and a prophage
embedded in the host genome \cite{Hendrix1999,Brussow2002,Lawrence2001}. Some
have argued that the latter mechanism occurs more frequently, due to the large
number of lysogenized prophages in bacterial genomes \cite{Lawrence2001}.

Lateral gene transfers could affect the codon usage patterns of phages,
especially if recombination occurs between phages whose preferred hosts
differ. In this case, the codon usage patterns of each phage may be expected to
reflect the preferred codons of their preferred hosts; a recent recombination
may result in regions of dramatically different codon usage from the average
phage codon usage. In particular, regions of unusual GC3 content in a phage
genome could reflect gene transfers between phages that typically infect hosts
of different GC3 content, in analogy with lateral gene transfer amongst
bacteria \cite{Ochman2000}. Morons are genes in phage genomes that are under
different transcriptional control than the rest of the phage genes, and are
often expressed when the phage is in the lysogenic state \cite{Hendrix2000}.
These morons have been observed to have very different nucleotide compositions
compared to the rest of the phage genome suggesting that they are the result of
such gene transfers \cite{Hendrix2000}. Thus one interpretation for our
observations of the 29 phages exhibiting non-random GC3 patterns is that these
genomes arose through recent recombination events, and have not subsequently
experienced enough time to equilibrate their GC3 content to that of their
current host. Given the lack of reliable estimates for time scales between
putative phage recombination events, or for codon usage equilibration, this
study neither supports nor refutes this interpretation. However, the
predominance of significant non-random patterns of GC3 in the genomes of
temperate phages (see Table \ref{tab:temperate_non}) may suggest that such
recombination occurs more frequently among temperate phage populations.

We have demonstrated that phage genes encoding structural proteins exhibit
significantly elevated CAI values compared the non-structural phage genes. These
results support the classical translation selection hypothesis, now extended to
the relationship between viral and host codon usage. We do not find much
variation in codon usage among the structural genes themselves. This observation
has two plausible interpretations within the literature of lateral gene
transfers: either phages of different preferred hosts rarely co-infect, or there
is substantially less recombination among the structural proteins of phages. The
latter hypothesis has been independently suggested for the capsid proteins of
phages, based on the idea that capsid proteins form a complex with
multiple physical interactions whose function would be disrupted by individual
gene transfer events \cite{Hendrix2002}. Unlike capsid genes, phage tail genes
often exhibit mosaicism, and they they can include elements from diverse viruses
with variable host ranges \cite{Haggard-Ljungquist1992,Hendrix2002}. To
investigate this phenomenon in the context of codon usage, we refined the
structural annotation to separate head from tail genes (see Section
Methods). We performed three separate ANOVAs to compare
the CAI usage in these genes: comparing head versus non-structural, tail versus
non-structural, and head versus tail (Table
\ref{tab:all_head_tail_aqua_orange}). These regressions indicate that the head
genes are primarily responsible for that pattern of elevated CAI in structural
proteins. In addition, we detect a difference in codon usage between head and
tail genes. These results have at least two possible explanations: either the
head proteins are produced in higher copy number than the tail proteins, or
lateral gene transfers between diverse phages occur frequently enough in the
tail genes to impair their ability to optimize codon usage to their current
host. The first hypothesis is very plausible, in light of evidence on the copy
number of head and tail proteins \cite{Hendrix2004}; nevertheless, we cannot
rule out the second possibility.

\section{Materials and Methods}\label{sec:materials_and_methods} 

\subsection{Bacteriophage Genomes}\label{sub:bacteriophage_genomes} 

Bacteriophage genomes were downloaded from NCBI's GenBank
(\verb=http://www.ncbi.nlm.nih.gov/Genbank/index.html=) release 156 (October,
2006) using Biopython's \cite{biopython} NCBI interface. We only used
reference sequence (refseq) phage genome records with accessions
of the form NC\_00dddd in order to have the most complete records
available. Of the 396 phage refseq's available, we focused on the 74 genomes of
phages whose primary host, as listed in the \verb=specific_host= tag in the
GenBank file, were \emph{E. coli}, \emph{P. aeruginosa} or \emph{L. lactis}. (A
complete list of the accession numbers used can be found in the supplementary
material.)

All phage genomes were downloaded from GenBank. Before being used for the rest
of this study, every gene within a genome was scanned for overlaps within other
genes in the same genome, and all overlapping sequences were removed. A codon
was only retained if all three of its nucleotides occurred in a single open
reading frame. Thus the final genome sequence used was a concatenation
of all non-overlapping coding sequences, omitting any control elements and other
non-coding sequences.

\subsection{Calculation of CAI Master
Tables}\label{sub:calculation_of_cai_master_tables} 

The definition of the Codon Adaptation Index requires the construction of a
`master' $w$-table for the host organism. Each of the 61 sense codons is
assigned a $w$-value based on the codon's frequency among the most highly
expressed genes in the host organism. In defining this set of genes, we follow
Sharp \cite{Sharp1987}, who specified highly expressed genes for \emph{E. coli}.

In order to calculate the CAI master $w$-tables for P. aeruginosa and L. lactis,
we identified the homologs of the highly expressed \emph{E. coli} genes within
the other host genomes, using BLAST \cite{Altschul1990}. In particular, we used
qblast to find homologs to these \emph{E. coli} genes by inputting the gene
protein sequences, and blasting (blastp) against the nr database, restricting
the database to include proteins of the target organism. In all cases, we used
the most significant blast result as the ortholog, provided its e-value was less
than $1\mathrm{x}10^{-10}$.

The particular proteins used for each of these three hosts are as
follows (NCBI genome accession numbers listed in parentheses beside the
host name, gI numbers listed in parentheses beside each protein). \emph{E.
coli} (NC\_000913): 30S ribosomal protein S10 (16131200), 30S ribosomal
protein S21 (16130961), 30S ribosomal protein S12 (16131221), 30S
ribosomal protein S20 (16128017), 30S ribosomal protein S1 (16128878),
30S ribosomal protein S2 (16128162), 30S ribosomal protein S15
(16131057), 30S ribosomal protein S7 (16131220), 50S ribosomal protein
L28 (16131508), 50S ribosomal protein L33 (16131507), 50S ribosomal
protein L34 (16131571), 50S ribosomal protein L11 (16131813), 50S
ribosomal protein L10 (16131815), 50S ribosomal protein L1 (1790416 ),
50S ribosomal protein L7/L12 (1790418 ), 50S ribosomal protein L17
(16131173), 50S ribosomal protein L3 (16131199), murein lipoprotein
(16129633), outer membrane protein A (3a;II*;G;d) (16128924), outer
membrane porin protein C (16130152), outer membrane porin 1a (Ia;b;F)
(16128896), protein chain elongation factor EF-Tu (duplicate of tufB)
(16131218), TufB (29140507), elongation factor Ts (16128163), elongation
factor EF-2 (16131219), recombinase A (16130606), molecular chaperone
DnaK (16128008); \emph{P. aeruginosa} (NC\_002516): elongation factor G
(15599462), 30S ribosomal protein S10 (15599460), 30S ribosomal protein
S21 (15595776), 30S ribosomal protein S12 (15599464), 30S ribosomal
protein S20 (15599759), 30S ribosomal protein S1 (15598358), 30S
ribosomal protein S2 (15598852), 30S ribosomal protein S15 (15599935),
30S ribosomal protein S7 (15599463), 50S ribosomal protein L28
(15600509), 50S ribosomal protein L33 (15600508), 50S ribosomal protein
L34 (15600763), 50S ribosomal protein L11 (15599470), 50S ribosomal
protein L10 (15599468), 50S ribosomal protein L1 (15599469), 50S
ribosomal protein L7/L12 (15599467), 50S ribosomal protein L17
(15599433), 50S ribosomal protein L3 (15599459), probable outer membrane
protein precursor (15596238), elongation factor Tu (15599461),
elongation factor Ts (15598851), elongation factor G (15599462),
recombinase A (15598813), molecular chaperone DnaK (15599955); \emph{L. lactis}
(NC\_002662): 30S ribosomal protein S10 (15674082), 30S ribosomal
protein S21 (15672222), 30S ribosomal protein S12 (15674244), 30S
ribosomal protein S20 (15673721), 30S ribosomal protein S1 (15672820),
30S ribosomal protein S2 (15674135), 30S ribosomal protein S15
(15673868), 30S ribosomal protein S7 (15674243), 50S ribosomal protein
L34 (15672113), 50S ribosomal protein L11 (15673983), 50S ribosomal
protein L10 (15673251), 50S ribosomal protein L1 (15673982), 50S
ribosomal protein L7/L12 (15673250), 50S ribosomal protein L17
(15674049), 50S ribosomal protein L3 (15674081), elongation factor Tu
(15673843), elongation factor Ts (15674134), elongation factor EF-2
(15674242), recombinase A (15672336), molecular chaperone DnaK
(15672936).

Given the set of highly expressed genes, the CAI master $w$-table was
calculated as follows. For each host, the GenBank file (GenBank release
156) was downloaded locally and transformed into a local data
structure using Biopython's \cite{biopython} GenBank parser. The
data structure was then scanned for each of the genes in the
highly translated gene set, and the collective CDS codon sequences of
these genes were concatenated together into one long sequence. Stop
codons and codons encoding for amino acids methionine (M), and
tryptophan (W) (each encoded by only one codon) were removed
from the concatened sequence. The frequencies of codons encoding all
other amino acids were then tabulated, and divided into groups according
to which amino acid they encode. The w-values are then calculated,
according to the procedure of Sharp \cite{Sharp1987}, as these
frequencies, normalized by the maximum frequency within each group. Thus
each amino acid has a codon with a $w$-value of 1, representing
the most commonly used codon for that amino acid. The $w$-values for
the stop codons and codons for methionine and tryptophan were set to the
average w-value of the remaining codons.

\subsection{Drawing Random Genomes According to
Constraints}\label{sub:drawing_random_genomes_according_to_constraints} 

Our randomization tests require drawing randomized phage genomes that are
constrained to have specific properties. In all of the randomization tests
discussed, the random sequences were drawn as a sequence of synonymous codons at
each position, thereby exactly preserving the amino acid sequences of proteins.

The three randomization tests used in this work can all be considered variants
of a canonical randomization test that preserves both the amino acid sequence
and a bit mask sequence exactly, while drawing codons from the global,
genome-wide distribution. A bit mask sequence is string of zeros and ones
corresponding to all codons in the genome. For example, GC3 is 1 if the third
position of a codon is G or C, and 0 otherwise.

Using the GC3 bit mask as an example, the randomization test procedure is
initialized by calculating the global codon frequencies that fit into categories
specified by the amino acid and the bit-mask value. Each amino acid has
associated with it two distributions: one for a bit-mask value of 1 and one for
a bit-mask value of 0. For example, alanine (A), is encoded by four codons, GCC
(1), GCG (1), GCT (0), GCA (0), where the GC3 bit-mask is shown in parenthesis.
Thus to calculate the codon distribution of alanine GC3 codons ($A_1$), we
compute the frequency of GCC and GCG codons across the whole phage genome.
Similarly, the distribution of $A_0$ codons is determined from the frequency of
GCT and GCA codons across the genome. In order to produce a random genome,
random codons are drawn at each position according to the distribution
associated with the position's amino acid and bit-mask value.

Thus the three null tests can be specified by the definition of the bit mask
along the sequence, which determines the constraints on the
randomize trials. The aqua randomization test constrains the amino acid
sequence and nothing else, and so its bit mask consists of all 1's. The orange
randomization test preserves the amino acid and the GC3, and so its bit mask is
the GC3 sequence mentioned above. The green randomization test preserves the
amino acid and BCAI exactly, thus its bit mask is the thresholded BCAI (1 if
BCAI $\geq$ 0.7, 0 otherwise).

\subsection{Structural Annotation}\label{sub:structural_annotation} 

All phage genes were annotated as structural or non-structural by inspecting
the annotations of high-scoring BLAST hits among viral proteins. This procedure is
described in detail below.

Each gene was considered separately within each genome object, although overlaps
were removed in the process of creating the genome objects (see section
\ref{sub:bacteriophage_genomes}). The amino acid sequence of each gene was
blasted against all known viral protein sequences using Biopython's interface
\cite{biopython} to the NCBI blast utility \cite{Altschul1990}. Specifically, we
used the blastp utility specifying the nr database, with entrez query `Viruses
[ORGN]'. We retained only those BLAST hits with e-values below the cutoff
$1\mathrm{x}10^{-4}$. All words in the title of these BLAST hits were collected,
using white space as a word-delimiter.

The unique words from the blast hits were then compared against a set of
structural keywords: ``capsid", ``structural", ``head", ``tail", ``fiber",
``scaffold", ``portal", ``coat", and ``tape". The words associated with the
BLAST hits were scanned for matches to the keywords, where each keyword was
treated as a regular expression. As a result, partial matching was counted as a
match. For example, a BLAST title containing the word `head-tail' would match
both keywords `head' and `tail'. If a gene had at least one structural keyword
match in its BLAST hit title, it was annotated as structural. Otherwise, it was
annotated as non-structural.

We further subdivided the structural annotation into two classes: head and tail
genes. Tail genes were identified with the keywords ``tail", ``fiber", and
``tape". These remaining structural genes that did not contain any of these
keywords were annotated as head genes. Two false positives for tail
identification in the lambda phage genome were manually corrected.

\subsection{Null Model: Results for Random Walk
Landscapes}\label{sub:null_model_results_for_random_walk_landscapes} 


In the sections above we have compared the genome landscapes calculated
from real genome sequences to a null model in which the sequences are
randomly drawn from a defined distribution. In this section, we compute
several properties of genome landscapes calculated from these random
genomes.

We write the general genome landscape of length $N$ as
\begin{equation}
    F(m) = \sum_{i=1}^m (\eta(i) - \overline{\eta}),
\end{equation}
where $\eta(i)$ are indepedant, and chosen from a random distribution with
$\mathrm{var}(\eta(i)) = \langle \eta(i)^2 \rangle - \langle \eta(i) \rangle^2 =
\Delta$, and
\begin{equation}
    \overline{\eta} = \frac{1}{N}\sum_{i=1}^N \eta(i),
\end{equation}
which ensures $F(0) = F(N) = 0$.

The purple regions in Figure \ref{fig:land_hist} represent the variance in the
genome landscapes of this null model at each $m$, $\sigma(m) = \sqrt{\langle F(m)^2
\rangle - \langle F(m) \rangle^2}$. Using the definitions above, we have
\begin{equation}
    \begin{aligned}
        F(m) &= \sum_{i=1}^m \eta(i)- \frac{m}{N}\sum_{i=1}^N \eta(i) \\
             &= \left( \frac{m + (N-m)}{N} \right) \sum_{i=1}^m \eta(i)- \frac{m}{N}\sum_{i=1}^N \eta(i) \\
             &= \frac{N-m}{N}\sum_{i=1}^m \eta(i) - \frac{m}{N}\sum_{i=m+1}^N \eta(i),
    \end{aligned}
\end{equation}
and
\begin{equation}
    \langle F(m) \rangle = \frac{m(N-m)\langle\eta\rangle}{N} - \frac{m(N-m)\langle\eta\rangle}{N} = 0.
\end{equation}
When we use $\langle \eta(i)\eta(j) \rangle = \langle \eta^2 \rangle \delta_{i,j}
+ (1- \delta_{i,j}) \langle\eta\rangle^2$, with $\delta_{i,j} = 1$ if $i = j$ and 0 otherwise, we find
\begin{equation}
    \begin{aligned}
    \langle F(m)^2 \rangle &= \frac{m(N-m)}{N} (\langle\eta^2\rangle - \langle\eta\rangle^2) \\
    &= \frac{\Delta m(N-m)}{N},
    \end{aligned}
\end{equation}
leading to $\sigma(m) = \sqrt{\langle F(m)^2 \rangle - \langle F(m) \rangle^2} = \sqrt{\Delta m
(N-m)/N}$. In the case of GC3 landscapes, $\eta(i)$ is either 1 or 0 with equal
probability, giving $\Delta_{\mathrm{GC3}} = 1/4$.

We can also calculate the full probability distribution,
$P(f;m,N,\Delta)$ that the genome landscape of length $N$ has an intermediate
value $F(m) = f$, at point $m$, by considering an $N$-step random walk that is
constrained to start and stop at $0$. This probability distribution can be
written as a product of two conditional probabilities for a walk that starts at
$0$ and ends at $f$ in $m$ steps, and a walk that starts at $f$ and ends at $0$
in $N-m$ steps
\begin{eqnarray}
    \begin{aligned}
        \label{eq:P_decomp}
    P(f;m,N,\Delta) &= A G(0,f;m,\Delta) G(f,0;N-m,\Delta) \\
                    &= A G(0,f;m,\Delta) G(0,f;N-m,\Delta),
    \end{aligned}
\end{eqnarray}
where $A$ is a normalization constant, and the last step used the inversion
symmetry of the random walks. Thus we seek the form of the conditional
probability $G(0,f;m,\Delta)$. In the same way as in Eq. (\ref{eq:P_decomp}), we
decompose this conditional probability into a multiplication of the conditional
probabilities for two walks, one that starts at $0$ and ends at $y$ in $x$
steps, and one that starts at $y$ and ends at $f$ in $m-x$ steps, and integrate
over all possible intermediate values $y$
\begin{equation}
    G(0,f;m,\Delta) = \int_{-\infty}^{\infty} \mathrm{d}y G(0,y;x,\Delta) G(y,f;m-x,\Delta).
\end{equation}
We can continue this decomposition for each intermediate step to give
\begin{equation}
    G(0,f;m,\Delta) = \int_{-\infty}^{\infty} \mathrm{d}y_1 \ldots \int_{-\infty}^{\infty} \mathrm{d}y_{m-1} G(0,y_1;1,\Delta) G(y_1,y_2;1,\Delta) \ldots G(y_{m-1},f;1,\Delta).
\end{equation}
Keeping the order of integration the same, and noting that $G(y_1,y_2;1,\Delta)
= G(y_2 - y_1;1,\Delta)$ for these random walks, we can write $y_{i+1} - y_i =
s_{i+1}$ to give
\begin{equation}
    G(0,f;m,\Delta) = \int_{-\infty}^{\infty} \mathrm{d}s_1 \ldots \int_{-\infty}^{\infty} \mathrm{d}s_m G(s_1;1,\Delta) \ldots G(s_2;1,\Delta) G(s_m;1,\Delta) \delta\left( \sum_{i=1}^m s_m - f\right),
\end{equation}
where the delta function is added to force the constraint that the sum of all
the intermediate steps must be equal to $f$. All of the intermediate conditional
probabilities now represent one step walks, and so are equal to the underlying
probability distribution of drawing a step size $s_m$, $p(s_m;\Delta)$
\begin{equation}
    G(0,f;m,\Delta) = \int_{-\infty}^{\infty} \mathrm{d}s_1 \ldots \int_{-\infty}^{\infty} \mathrm{d}s_m \delta\left( \sum_{i=1}^m s_m - f\right) \Pi_{i=1}^m p(s_i;\Delta).
\end{equation}
Making use of the integral representation of the delta function \cite{Grosberg1994}
\begin{equation}
    \delta(x) = \frac{1}{2\pi} \int_{-\infty}^{\infty} \mathrm{d}k e^{-ikx},
\end{equation}
we have
\begin{equation}
    G(0,f;m,\Delta) = \frac{1}{2\pi} \int_{-\infty}^{\infty} \mathrm{d}k e^{-ikf} \tilde{p}(k;\Delta)^m,
\end{equation}
where $\tilde{p}(k;\Delta)$ is the Fourier transform of $p(s;\Delta)$
\begin{equation}
    \tilde{p}(k;\Delta) = \int_{-\infty}^{\infty} \mathrm{d}s e^{-iks} p(s;\Delta) .
\end{equation}
For the purpose of this discussion, we assume $p(s;\Delta)$ has a Gaussian form
$p(s) = \frac{1}{\sqrt{2\pi\Delta}}e^{-\frac{s^2}{2\Delta}}$, and note that the
results are general. In this case, $\tilde{p}(k;\Delta) =
e^{-\frac{k^2\Delta}{2}}$, and we have
\begin{equation}
    G(0,f;m) = \frac{1}{2\pi} \int_{-\infty}^{\infty} \mathrm{d}k e^{-m\Delta k^2/2}e^{-ikf} = \frac{1}{\sqrt{2\pi m\Delta}}e^{-f^2/2m\Delta}.
\end{equation}
To determine $A$, we enforce the normalization condition
\begin{equation}
    \int_{-\infty}^{\infty} \mathrm{d}f P(f;m,N,\Delta) = 1,
\end{equation}
which gives
\begin{eqnarray}
    \begin{aligned}
    P(f;m,N,\Delta) &= \frac{1}{\sigma\sqrt{2\pi}}e^{-f^2/2\sigma^2} \\
    \sigma(m) &= \sqrt{\Delta\frac{m(N-m)}{N}}.
    \end{aligned}
\end{eqnarray}
Note that from the full distribution, we can immediately identify $\sigma(m) =
\sqrt{\langle F(m)^2 \rangle - \langle F(m) \rangle^2}$, confirming the explicit
calculation above.

\subsection{Acknowledgments}\label{sub:acknowledgments}

The authors would like to thank Herv\'{e} Isambert, Graham Hatfull, and Roger
Hendrix for conversations and suggestions on this work. JBL and DRN would like
to thank the Institute Curie, Paris, for hospitality during the initial phases
of this work. Work by DRN was supported by the National Science Foundation
through grants DMR-0231631 and DMR-0213805. JBL acknowledges the financial
support of the Fannie and John Hertz Foundation. JBP acknowledges
support from the Burroughs Wellcome Fund.



\clearpage
\clearpage

\clearpage
\clearpage
\begin{figure}
	[p] 
	\begin{center}
		\includegraphics[scale=0.6]{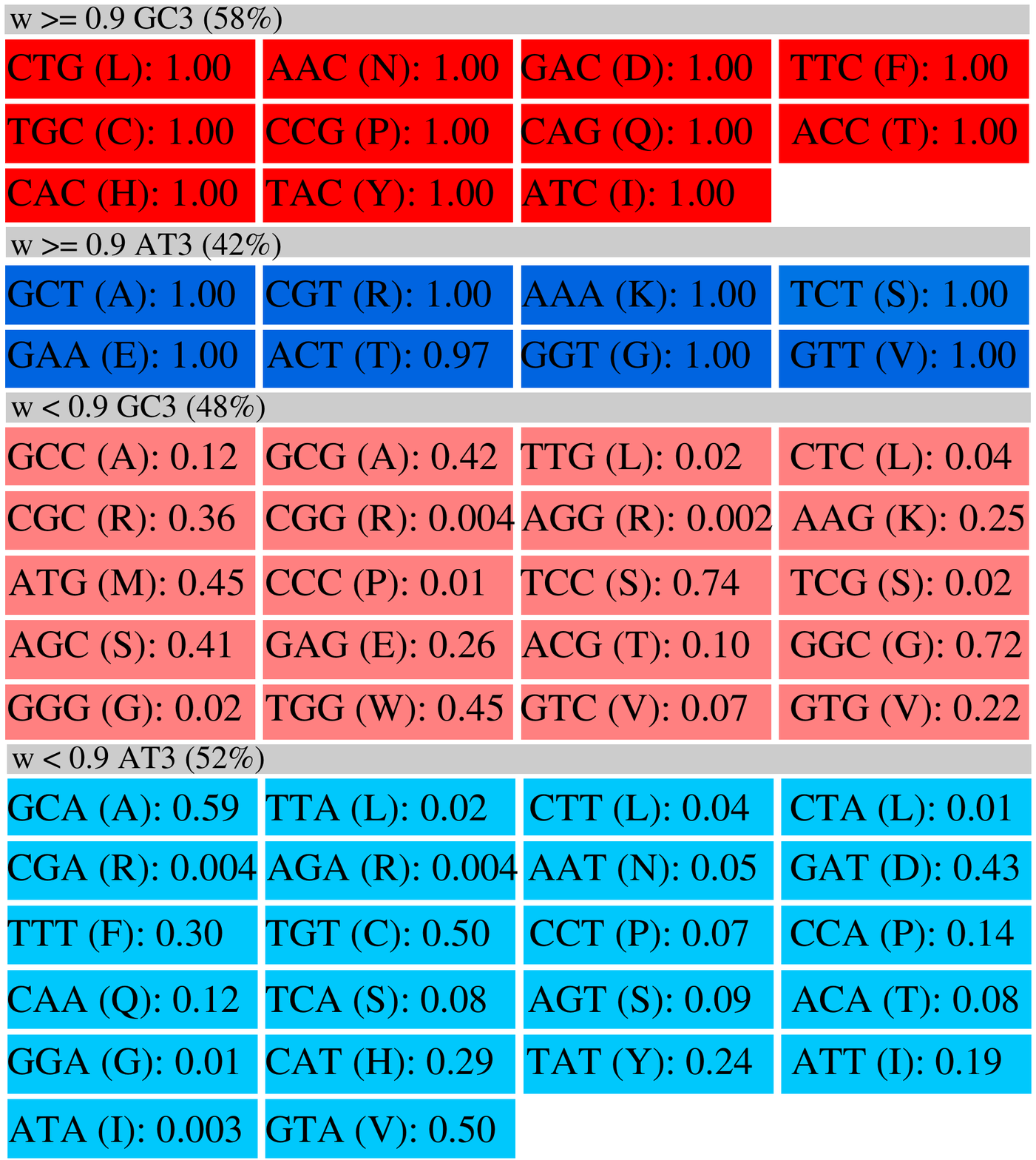} 
	\end{center}
	\caption{{\bf \emph{E. coli} codon usage master table.} The table of 61 codons 
	along with their associated w-values is shown for \emph{E. coli}. 
	The $w$-value of each codon reflects its frequency in 
	highly transcribed \emph{E. coli} genes (see main text). The table 
	is divided into four regions: codons with high CAI ($w \geq 0.9$) 
	ending in G or C (dark red); codons with high CAI ending in A or 
	T (dark blue); codons with low CAI ($w \leq 0.9$) ending in G or C 
	(light red); codons with low CAI ending in A or T (light blue). 
	As the table shows, there is 
	a slight bias for GC3 in the high-CAI codons (58\%), and slight 
	bias away from GC3 in the low-CAI codons (48\%).} \label{fig:E_coli_master} 
\end{figure}
\clearpage
\clearpage
\begin{figure}
	[p] 
	\begin{center}
		\begin{tabular}
			{ccc} 
			\includegraphics[scale=1]{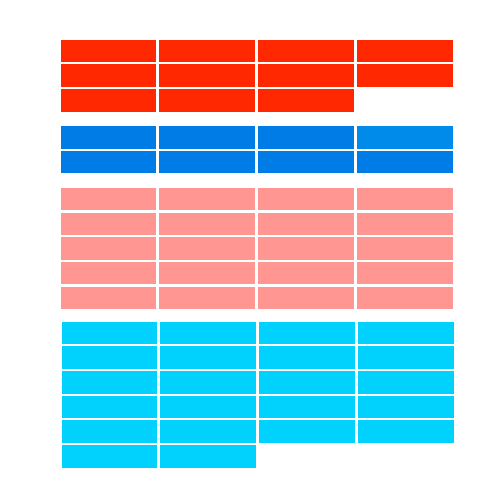} & 
			\includegraphics[scale=1]{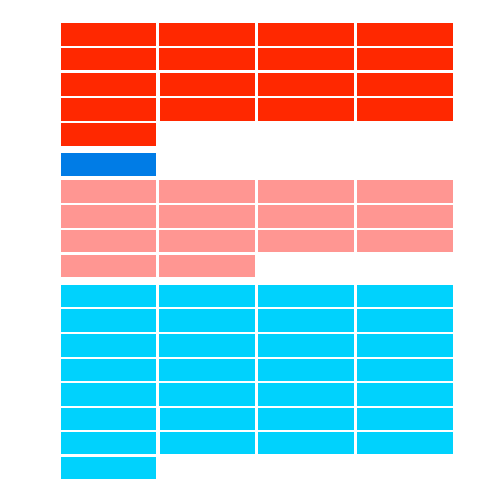} & 
			\includegraphics[scale=1]{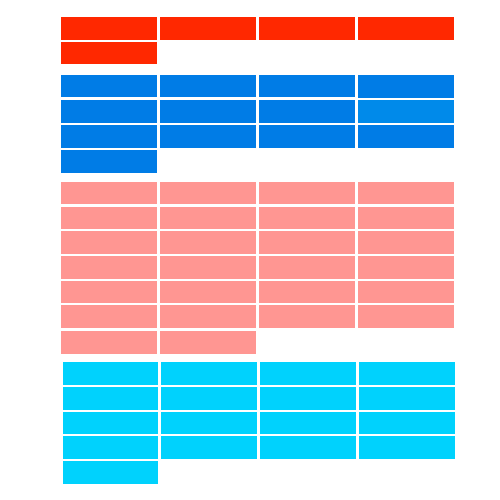} \\
		\end{tabular}
	\end{center}
	\caption{{\bf Schematics of prefered codon usage tables for 
	\emph{E. coli}, \emph{P. aeruginosa}, and \emph{L. lactis} following the 
	conventions of Figure \ref{fig:E_coli_master}.} 
	Unlike \emph{E. coli},
	\emph{P. aeruginosa} strongly favors GC3 in high-CAI codons 
	(94\%), and \emph{L. lactis} strongly favors AT3 in high-CAI 
	codons (72\%).} \label{fig:master_cartoons} 
\end{figure}
\clearpage
\clearpage
\clearpage
\clearpage


\begin{table}
    \begin{center}
    \begin{tabular}{c|c|c|c}
        Test Name & Genome Properties Constrained & Genome Properties Varied & Figure \\
        \hline
        Aqua & amino acid sequence, global codon distribution & synonymous codons & \ref{fig:aqua} \\
        Orange & amino acid and BCAI sequences & GC3 & \ref{fig:green_orange} \\
        Green & amino acid and GC3 sequences & BCAI & \ref{fig:green_orange} \\
    \end{tabular}
    \end{center}
    \caption{Randomization test descriptions.  
	 The three randomization tests used in the paper 
	 are color-coded according to what genome properties 
	 are constrained in the random trials.}
    \label{tab:tests}
\end{table}
\clearpage

\begin{table}
    {\tiny
    \begin{tabular}{c|c|c|c|c|c|c|c|c|c}
        Name & Host & Accession & Lifestyle & \# Genes & 
           Length & Coding Length & \%GC3 & Orange p-value & Green p-value \\
           \hline
           T5 & \ecoli & NC\_005859 & NT  & 161 & 121,750 & 96,051 & 31.6 & $1.38\mathrm{x}10^{-31}$ & $1.71\mathrm{x}10^{-19}$ \\
           RB69 & \ecoli & NC\_004928 &  NT & 273 & 167,560 & 156,147 & 29.0 & $1.25\mathrm{x}10^{-21}$ & $5.21\mathrm{x}10^{-01}$ \\
           phiEL & \paeru & NC\_007623 & NT  & 201 & 211,215 & 194,850 & 57.8 & $7.38\mathrm{x}10^{-20}$ & $2.17\mathrm{x}10^{-09}$ \\
           RB49 & \ecoli & NC\_005066 &NT   & 273 & 164,018 & 152,592 & 36.9 & $2.01\mathrm{x}10^{-18}$ & $2.48\mathrm{x}10^{-01}$ \\
           F116 & \paeru & NC\_006552 &  T & 70 & 65,195 & 60,240 & 76.3 & $1.31\mathrm{x}10^{-10}$ & $6.31\mathrm{x}10^{-16}$ \\
           CTX & \paeru & NC\_003278 &T   & 47 & 35,580 & 31,971 & 81.2 & $1.44\mathrm{x}10^{-09}$ & $6.82\mathrm{x}10^{-32}$ \\
           phiKMV & \paeru & NC\_005045 & NT  & 49 & 42,519 & 38,310 & 79.9 & $3.25\mathrm{x}10^{-09}$ & $9.54\mathrm{x}10^{-03}$ \\
           T4 & \ecoli & NC\_000866 &  NT & 269 & 168,903 & 153,660 & 24.3 & $4.59\mathrm{x}10^{-09}$ & $8.62\mathrm{x}10^{-01}$ \\
           lambda & \ecoli & NC\_001416 & T  & 69 & 48,502 & 40,773 & 53.5 & $6.25\mathrm{x}10^{-09}$ & $5.10\mathrm{x}10^{-68}$ \\
           D3 & \paeru & NC\_002484 & T  & 94 & 56,425 & 49,095 & 68.3 & $1.57\mathrm{x}10^{-08}$ & $3.85\mathrm{x}10^{-07}$ \\
           P2 & \ecoli & NC\_001895 & T  & 42 & 33,593 & 30,411 & 54.7 & $5.60\mathrm{x}10^{-08}$ & $2.54\mathrm{x}10^{-61}$ \\
           P1 & \ecoli & NC\_005856 & T  & 108 & 94,800 & 80,103 & 48.2 & $9.37\mathrm{x}10^{-08}$ & $3.51\mathrm{x}10^{-11}$ \\
           D3112 & \paeru & NC\_005178 & T  & 55 & 37,611 & 34,908 & 80.4 & $3.05\mathrm{x}10^{-07}$ & $4.35\mathrm{x}10^{-05}$ \\
           WPhi & \ecoli & NC\_005056 &T   & 43 & 32,684 & 29,601 & 56.4 & $8.39\mathrm{x}10^{-07}$ & $7.80\mathrm{x}10^{-55}$ \\
           K1F & \ecoli & NC\_007456 & NT  & 43 & 39,704 & 34,629 & 53.4 & $1.75\mathrm{x}10^{-05}$ & $8.03\mathrm{x}10^{-02}$ \\
           T3 & \ecoli & NC\_003298 &  NT & 47 & 38,208 & 29,694 & 54.3 & $3.50\mathrm{x}10^{-05}$ & $3.07\mathrm{x}10^{-04}$ \\
           PaP3 & \paeru & NC\_004466 &  T & 71 & 45,503 & 41,115 & 58.1 & $5.09\mathrm{x}10^{-05}$ & $1.64\mathrm{x}10^{-19}$ \\
           phiV10 & \ecoli & NC\_007804 & T  & 55 & 39,104 & 36,111 & 48.8 & $1.25\mathrm{x}10^{-04}$ & $9.38\mathrm{x}10^{-11}$ \\
           P27 & \ecoli & NC\_003356 &   T& 58 & 42,575 & 37,707 & 50.5 & $2.24\mathrm{x}10^{-04}$ & $2.23\mathrm{x}10^{-20}$ \\
           933W & \ecoli & NC\_000924 &  T & 78 & 61,670 & 52,956 & 50.0 & $4.29\mathrm{x}10^{-04}$ & $8.88\mathrm{x}10^{-09}$ \\
           B3 & \paeru & NC\_006548 &  T & 56 & 38,439 & 36,138 & 77.3 & $4.40\mathrm{x}10^{-04}$ & $3.33\mathrm{x}10^{-05}$ \\
           HK97 & \ecoli & NC\_002167 & T  & 59 & 39,732 & 34,191 & 52.1 & $7.61\mathrm{x}10^{-04}$ & $1.19\mathrm{x}10^{-20}$ \\
           VT2-Sa & \ecoli & NC\_000902 & T  & 83 & 60,942 & 52,647 & 51.3 & $1.31\mathrm{x}10^{-03}$ & $7.40\mathrm{x}10^{-07}$ \\
           PRD1 & \ecoli & NC\_001421 &  NT & 21 & 14,925 & 11,988 & 47.6 & $2.99\mathrm{x}10^{-03}$ & $5.97\mathrm{x}10^{-02}$ \\
           JK06 & \ecoli & NC\_007291 &  U & 71 & 46,072 & 32,841 & 43.0 & $3.84\mathrm{x}10^{-03}$ & $1.63\mathrm{x}10^{-03}$ \\
           T1 & \ecoli & NC\_005833 & NT  & 77 & 48,836 & 44,010 & 47.7 & $7.45\mathrm{x}10^{-03}$ & $3.64\mathrm{x}10^{-01}$ \\
           Pf1 & \paeru & NC\_001331 &  U & 12 & 7,349 & 6,282 & 75.7 & $9.66\mathrm{x}10^{-03}$ & $6.67\mathrm{x}10^{-01}$ \\
           HK022 & \ecoli & NC\_002166 & T  & 57 & 40,751 & 33,885 & 52.7 & $1.25\mathrm{x}10^{-02}$ & $4.36\mathrm{x}10^{-18}$ \\
           4268 & \llact & NC\_004746 &  NT & 49 & 36,596 & 33,759 & 24.7 & $1.59\mathrm{x}10^{-02}$ & $3.20\mathrm{x}10^{-01}$ \\
           BP-4795 & \ecoli & NC\_004813 & T  & 48 & 57,930 & 22,356 & 48.1 & $1.66\mathrm{x}10^{-02}$ & $3.29\mathrm{x}10^{-10}$ \\
           186 & \ecoli & NC\_001317 &T   & 43 & 30,624 & 27,747 & 58.7 & $4.02\mathrm{x}10^{-02}$ & $1.79\mathrm{x}10^{-22}$ \\
           I2-2 & \ecoli & NC\_001332 &  U & 8 & 6,744 & 5,166 & 35.0 & $6.91\mathrm{x}10^{-02}$ & $1.01\mathrm{x}10^{-01}$ \\
           phiKZ & \paeru & NC\_004629 & NT  & 306 & 280,334 & 243,384 & 26.8 & $1.32\mathrm{x}10^{-01}$ & $1.79\mathrm{x}10^{-14}$ \\
           bIL312 & \llact & NC\_002671 &  T & 27 & 15,179 & 11,292 & 28.1 & $1.49\mathrm{x}10^{-01}$ & $8.85\mathrm{x}10^{-04}$ \\
           HK620 & \ecoli & NC\_002730 &  T & 58 & 38,297 & 33,717 & 45.9 & $1.61\mathrm{x}10^{-01}$ & $1.41\mathrm{x}10^{-05}$ \\
           Mu & \ecoli & NC\_000929 & T  & 54 & 36,717 & 33,900 & 54.1 & $1.68\mathrm{x}10^{-01}$ & $4.49\mathrm{x}10^{-10}$ \\
           P4 & \ecoli & NC\_001609 &  T & 14 & 11,624 & 9,765 & 52.4 & $1.71\mathrm{x}10^{-01}$ & $4.17\mathrm{x}10^{-18}$ \\
           N15 & \ecoli & NC\_001901 &  T & 59 & 46,375 & 41,472 & 54.9 & $2.17\mathrm{x}10^{-01}$ & $1.38\mathrm{x}10^{-09}$ \\
           Stx2 I & \ecoli & NC\_003525 & T  & 97 & 61,765 & 34,932 & 48.4 & $3.04\mathrm{x}10^{-01}$ & $4.23\mathrm{x}10^{-04}$ \\
           bIL286 & \llact & NC\_002667 &  T & 61 & 41,834 & 38,694 & 24.8 & $3.68\mathrm{x}10^{-01}$ & $1.17\mathrm{x}10^{-01}$ \\
           Tuc2009 & \llact & NC\_002703 &  T & 56 & 38,347 & 35,178 & 28.0 & $4.08\mathrm{x}10^{-01}$ & $1.81\mathrm{x}10^{-02}$ \\
           Stx2 II & \ecoli & NC\_004914 &T   & 99 & 62,706 & 34,755 & 50.1 & $5.85\mathrm{x}10^{-01}$ & $9.94\mathrm{x}10^{-03}$ \\
           BK5-T & \llact & NC\_002796 &  T & 52 & 40,003 & 33,267 & 24.0 & $5.91\mathrm{x}10^{-01}$ & $6.68\mathrm{x}10^{-01}$ \\
           Stx1 & \ecoli & NC\_004913 &  T & 93 & 59,866 & 33,444 & 49.5 & $6.75\mathrm{x}10^{-01}$ & $2.97\mathrm{x}10^{-03}$ \\
           LC3 & \llact & NC\_005822 &T   & 51 & 32,172 & 29,607 & 24.6 & $7.31\mathrm{x}10^{-01}$ & $4.90\mathrm{x}10^{-01}$ \\
           ul36 & \llact & NC\_004066 &  NT & 58 & 36,798 & 32,400 & 27.7 & $8.64\mathrm{x}10^{-01}$ & $4.66\mathrm{x}10^{-02}$ \\
           Pf3 & \paeru & NC\_001418 &U   & 9 & 5,833 & 5,487 & 35.9 & $8.70\mathrm{x}10^{-01}$ & $1.64\mathrm{x}10^{-06}$ \\
           bIL285 & \llact & NC\_002666 &T   & 62 & 35,538 & 32,646 & 26.7 & $9.20\mathrm{x}10^{-01}$ & $9.93\mathrm{x}10^{-01}$ \\
           r1t & \llact & NC\_004302 &T   & 50 & 33,350 & 30,315 & 25.4 & $9.53\mathrm{x}10^{-01}$ & $6.03\mathrm{x}10^{-01}$ \\
           bIL170 & \llact & NC\_001909 & T  & 63 & 31,754 & 27,663 & 27.1 & $9.91\mathrm{x}10^{-01}$ & $8.71\mathrm{x}10^{-01}$ \\
    \end{tabular}
    }
    \caption{Phage properties. Properties are listed for all phages included in
    Figure \ref{fig:green_orange_pass_genomes}, in the same order based on the
    orange p-value. Lifestyle annotations are T (temperate), NT (non-temperate),
    U (unknown). The coding length refers to the length of all coding sequences
    concatenated together (see Methods.}
    \label{tab:phage_properties}
\end{table}

\begin{table}
	\begin{center}
		\begin{tabular}
		    {c|c|c}
		     & Lambda & All Phage Genes \\
		     \hline
		     Number structural & 7 & 279 \\
		     Number non-structural & 18 & 1022 \\
		     \hline
		     \multicolumn{3}{c}{Aqua CAI Randomization Test} \\
		     \hline
		     median $p^{>}$ structural & $1.3\mathrm{x}10^{-4}$ & $8.0\mathrm{x}10^{-3}$ \\
		     median $p^{>}$ non-structural & 1.0 & 1.0 \\
		     ANOVA significance & $p=4.5\mathrm{x}10^{-5}$ & $p=4.7\mathrm{x}10^{-12}$ \\
		     \hline
		     \multicolumn{3}{c} {Orange BCAI Randomization Test} \\
		     \hline
		     median $p^{>}$ structural & $2.8\mathrm{x}10^{-2}$ & $2.0\mathrm{x}10^{-1}$ \\
		     median $p^{>}$ non-structural & 0.98 & 0.73 \\
		     ANOVA significance & $p=1.8\mathrm{x}10^{-4}$ & $p=1.6\mathrm{x}10^{-15}$ \\
		\end{tabular}
	\end{center}
	\caption{Structural annotation verses codon usage.  The table shows
	the median $p^>$ values amoung structural and non-structural genes,
	under the aqua and orange randomization tests. Small $p^>$ values indicate
	significantly elevated CAI, controlling for the amino acid sequence
	(aqua test) and the GC3 sequence (orange test). We also report the
	significance of non-parametic ANOVAs that compare median $p^>$-values between
	the structural and non-structural genes. Analyses are limited to 
	those genes that pass the aqua test, as described in the main text;
	similar results are found without this restriction.
} 
	\label{tab:lambda_all_struct_non_aqua_orange} 
\end{table}
\clearpage

\begin{table}
	\begin{center}
		\begin{tabular}
		    {c|c}
		     & All Phage Genes \\
		     \hline
		     Number `Head'  & 145 \\
		     Number `Tail'  & 134 \\
		     Number non-structural (NS) & 1022 \\
		     \hline
		     \multicolumn{2}{c}{Aqua CAI Randomization Test} \\
		     \hline
		     median $p^{>}$ head &  $2.0\mathrm{x}10^{-3}$ \\
		     median $p^{>}$ tail &  $2.0\mathrm{x}10^{-2}$ \\
		     median $p^{>}$ NS &  1.0 \\
		     ANOVA Head vs NS & $p=6.4\mathrm{x}10^{-19}$ \\
		     ANOVA Tail vs NS & $p=1.8\mathrm{x}10^{-1}$ \\
		     ANOVA Head vs Tail & $p=2.1\mathrm{x}10^{-8}$ \\
		     \hline
		     \multicolumn{2}{c} {Orange BCAI Randomization Test} \\
		     \hline
		     median $p^{>}$ head &  $7.0\mathrm{x}10^{-2}$ \\
		     median $p^{>}$ tail &  $4.3\mathrm{x}10^{-1}$ \\
		     median $p^{>}$ NS &  0.73 \\
		     ANOVA Head vs NS  & $p=4.2\mathrm{x}10^{-21}$ \\
		     ANOVA Tail vs NS  & $p=1.7\mathrm{x}10^{-2}$ \\
		     ANOVA Head vs Tail  & $p=6.0\mathrm{x}10^{-8}$ \\
		\end{tabular}
	\end{center}
	\caption{Comparison between codon usage and refined structural
	annotations.
	As in Table \ref{tab:lambda_all_struct_non_aqua_orange}, 
	we compare the median aqua and orange $p^>$ values among head genes, tail
	genes, and non-structural genes. We report the significance of
	pairwise non-parametric ANOVAs comparing head to non-structural, tail
	to non-structural, and head to tail genes.
	These analyses are limited to genes that pass the aqua test; 
	similar results are found without this
	restriction.
}
	\label{tab:all_head_tail_aqua_orange} 
\end{table}

\clearpage
\begin{table}
	\begin{center}
		\begin{tabular}
		    {c|c}
		     \multicolumn{2}{c}{Median $p_{\text{combined}}^{\mathrm{orange}}$} \\
		     \hline
		     Temperate & $1.4\mathrm{x}10^{-2}$ \\
		     Non-temperate & $2.6\mathrm{x}10^{-5}$ \\
		     Un-identified & $4\mathrm{x}10^{-2}$ \\
		     ANOVA significance & $p = 0.1$ \\
		     \hline
		     \multicolumn{2}{c}{Median $p_{\text{combined}}^{\mathrm{green}}$} \\
		     \hline
		     Temperate & $5.1\mathrm{x}10^{-9}$ \\
		     Non-temperate & $7.0\mathrm{x}10^{-2}$ \\
		     Un-identified & $5\mathrm{x}10^{-2}$ \\
		     ANOVA significance & $p = 0.009$ \\
		\end{tabular}
	\end{center}
\caption{{\bf Phage lifestyle versus codon usage}. The table shows the median
$p_{\text{combined}}^{\mathrm{orange}}$ and
$p_{\text{combined}}^{\mathrm{green}}$ values among phages classified as
temperate, non-temperate, or un-identified for all phages included in Figure
\ref{fig:green_orange_pass_genomes} and Table \ref{tab:phage_properties}. Small
median $p_{\text{combined}}^{\mathrm{orange}}$ values indicate that these phages
have significantly non-random (in either direction) BCAI, controlling for the
amino acid sequence and the GC3 sequence, while small median
$p_{\text{combined}}^{\mathrm{green}}$ values indicate that these phages have
significantly non-random (in either direction) GC3, controlling for the amino
acid sequence and the BCAI sequence. We also report the significance of
non-parametic ANOVAs that compare these medians between these groups of phages.
}
	\label{tab:temperate_non} 
\end{table}
\clearpage
\bibliography{GLA,GLA_lux}

\end{document}